\documentclass[fleqn,usenatbib]{mnras}

\usepackage{newtxtext,newtxmath}

\usepackage{xcolor}

\usepackage[T1]{fontenc}

\DeclareRobustCommand{\VAN}[3]{#2}
\let\VANthebibliography\thebibliography
\def\thebibliography{\DeclareRobustCommand{\VAN}[3]{##3}\VANthebibliography}

\usepackage{graphicx}	% Including figure files
\usepackage{amsmath}	% Advanced maths commands
\usepackage{subfig}   % For Subfigure
\title[Ionospheric effects on global 21cm ]{Extracting the Global 21-cm signal from Cosmic Dawn and Epoch of Reionization in the presence of Foreground and Ionosphere}

\author[Tripathi A. et al.]{
Anshuman Tripathi,$^{1}$\thanks{E-mail:anshumantripathi85@gmail.com}
Abhirup Datta,$^{1}$
Madhurima Choudhury$^{1,2,3}$,
Suman Majumdar$^{1,4}$
\\
$^{1}$Department of Astronomy, Astrophysics and Space Engineering, Indian Institute of Technology, Indore-453552, M.P., India\\
$^{2}$Astrophysics Research Center (ARCO), Department of Natural Sciences, The Open University of Israel, Ra’anana 4353701, Israel\\
$^{3}$ Department of physics, Brown University, Rhode Island 02914, USA \\
$^{4}$Department of physics, Blackett Laboratory, Imperial College, London SW7 2AZ, U. K.
}

\date{Accepted XXX. Received YYY; in original form ZZZ}

\pubyear{2022}

\begin{document}
\label{firstpage}
\pagerange{\pageref{firstpage}--\pageref{lastpage}}
\maketitle
\begin{abstract}
Detection of redshifted \ion{H}{i} 21-cm emission is a potential probe for investigating the Universe's first billion years. However, given the significantly brighter foreground, detecting 21-cm is observationally difficult. The Earth's ionosphere considerably distorts the signal at low frequencies by introducing directional-dependent effects. Here, for the first time, we report the use of Artificial Neural Networks (ANNs) to extract the global 21-cm signal characteristics from the composite all-sky averaged signal, including foreground and ionospheric effects such as refraction, absorption, and thermal emission from the ionosphere's F and D-layers. We assume a 'perfect' instrument and neglect instrumental calibration and beam effects. To model the ionospheric effect, we considered the static and time-varying ionospheric conditions for the mid-latitude region where LOFAR is situated. In this work, we trained the ANN model for various situations using a synthetic set of the global 21-cm signals created by altering its parameter space based on the "$\rm \tanh$" parametrized model and the Accelerated Reionization Era Simulations (ARES) algorithm. The obtained result shows that the ANN model can extract the global signal parameters with an accuracy of $\ge 96 \% $ in the final study when we include foreground and ionospheric effects. On the other hand, a similar ANN model can extract the signal parameters from the final prediction data set with an accuracy ranging from $97 \%$ to $98 \%$ when considering more realistic sets of the global 21-cm signals based on physical models.  
\end{abstract}


%%%%%%%%%%%%%%%%% BODY OF PAPER %%%%%%%%%%%%%%%%%%%%%%%%%%%%%%%%%%%%%%%%%%%%%%%%%%%%%%%%

\section{Introduction}
\label{ section 1}
The period from the beginning of star and galaxy formation [Cosmic Dawn (CD)] till the change of the state of the Universe from an absolutely neutral to a fully ionized state, i.e., the Epoch of Reionization (EoR), is still observationally unexplored to astronomers. Detection of the redshifted \ion{H}{i} 21-cm line is noticed as one of the most promising future probes of the Universe at these redshifts (z$\approx$ $7-30$) \citep[][]{furlanetto2006effects, pritchard201221}. The redshifted \ion{H}{i} 21-cm lines are formed due to hyperfine splitting of the 1S ground state. Studying these epochs can answer many essential cosmological queries, such as the features of the early galaxies, the physics of mini-quasars, the development of very metal-poor stars, and other major research topics on the origin and evolution of the Universe. The primary science goal of upcoming radio telescopes like the SKA is to study these three extended epochs of the universe's structure formation history. In past decades, significant progress has been made in the theoretical modelling of the expected redshifted 21-cm signal. There are two different experimental techniques for observing these signals in the observational field.
\citep{pritchard201221,harker2010power,shaver1999can}:
\newline (a) using several dishes and huge interferometric arrays at very low radio frequencies to obtain statistical power spectra of the \ion{H}{i} 21-cm variations, for example, Giant Meterwave Radio
Telescope (GMRT \citep{gmrt1991}), Hydrogen Epoch of Reionization Array (HERA \citep{HERA2017}), Low Frequency Array (LOFAR\citep{lofar2013}), Murchison Wide-field Array (MWA \citep{mwa2013}), Square Kilometer Array(SKA \citep{ska2015}), etc.
\newline(b) using a single radiometer to observe the sky-averaged signature of the redshifted \ion{H}{i} 21-cm line, for example, Broadband Instrument for global Hydrogen Reionization Signal (BIGHORN \citep{Bighorn2015}), Large-Aperture Experiment to Detect the Dark Ages (LEDA \citep{LEDA}), Experiment to Detect the global EoR Signature (EDGES \citep{bowman2018absorption}), Shaped Antenna measurement of the background Radio Spectrum (SARAS \citep{Patra2013,Singh2017}),etc.

Recently, the EDGES team announced a probable discovery of the Cosmic Dawn's sky-averaged \ion{H}{i} 21-cm global signal. They observed that the measured signal had an absorption trough double the magnitude expected by the standard cosmological model \citep{bowman2018absorption}. However, this supposed detection has been challenged by another independent experiment SARAS \citep{2022Nat_Saras}. This contradiction in independent detection of the global 21-cm signal from ground-based observation further highlights its challenges. One of the reason why this signal is very difficult to detect is because it is very faint. The signal is embedded behind a sea of dazzling galactic as well as extragalactic foregrounds. The magnitude of the foregrounds is several orders brighter than the signal, approximately $10^4$ to $10^6$ order brighter than the signal. Furthermore, human-made radio frequency interference (RFI), mainly by the FM band and Earth's ionosphere, will also affect ground-based observation. The ionosphere distorts the lower frequency signal significantly when it passes through the ionosphere.

The ionosphere is the uppermost layer of the atmosphere, extending from $\sim 50$ to $\sim 600 $ km above the Earth's surface. The impacts of solar activity significantly affect the electron density in the ionosphere. The ionospheric existence causes three significant effects in detecting the redshifted global 21-cm signal from the ground-based antenna. All radio waves, including galactic and extragalactic foregrounds, are refracted by the ionosphere, which also attenuates any trans-ionospheric signal and emits thermal radiation
\citep[][]{pawsey1951ionospheric,steiger1961observations}. Further, due to solar activation of the ionosphere, these effects are fundamentally time variable \citep[][]{evans1968radar,davies1990ionospheric}. These ionospheric effects scale as $\nu^{-2}$, where $\nu$ represents the frequency of observations. Hence, as the observing frequencies get lower, the effect of the ionosphere increases more. It demonstrates that when detecting the signal from the Cosmic Dawn and the Dark Ages ($z \ge 15$), ionospheric effects will have a stronger influence on global signals than when detecting the signal from the Epoch of Reionization ($15 \ge z \ge 6$) \citep{datta2016effects}.

The effects of static ionosphere refraction and absorption for ground-based observation between $30$ and $100$ MHz were previously examined by \citet{vedantham2014chromatic}. In \citet{datta2016effects}, they presented the dynamic ionosphere effects like refraction, absorption, and thermal emission. They also demonstrated how these combined effects affect the global 21-cm signal from Epoch Reionization and Cosmic Dawn when we are observing from the ground. \citet{emaa_2021, emma2022} recently investigated the chromatic ionospheric effects on global 21-cm data by modelling the two principal ionospheric layers, the F and D layers, with a reduced spatial model with temporal variance.
The investigation focuses on the chromatic distortions induced by the ionosphere.

Several studies have been done in recent years based on machine learning (ML) techniques to perform signal parameter estimation or signal modelling. \citet{shimabukuro2017analysing} and \citet{jennings2019evaluating} have employed machine learning techniques to predict 21-cm power spectrum parameters. Similarly, \citet{2022Madhurima_3rd} extends the ANN to extract the 21-cm PS and corresponding EoR parameters from synthetic observations for different telescope models. \citet{schmit2018emulation} used Artificial Neural Network (ANN) to emulate the 21-cm power spectrum for a wide range of parameters. Similarly, \citet{tiwari2022} have developed an ANN-based emulator for the signal bispectrum, which they have further used to estimate signal parameters via a Bayesian inference pipeline. The global 21-cm signal from Cosmic Dawn and EoR has also been emulated using ANN by \citet{cohen2020emulating, globalemu, VAE}. Convolutional Neural Networks (CNN) have been utilized to identify reionization sources from 21-cm maps \citet{hassan2019identifying}. \citet{chardin2019deep} and \citet{ korber22} have used deep learning models to emulate 21-cm maps from the dark matter distribution directly. \citet{gillet2019deep} used deep learning with CNN to predict astrophysical parameters directly from 21-cm maps. \citet{Zhao2022} used CNN to estimate parameters and infer posteriors on 3D-tomographic 21-cm images. An ANN model that can extract astrophysical parameters of 21 cm from mock observation data sets, including the effects of foregrounds, instruments, and noise, has been successfully developed and presented by \citet{choudhury2020extracting, choudhury2021using}.
 The relevance of non-parametric techniques for this purpose has already been demonstrated in several previous studies \citep{Harker2009, Tauscher2018}. These studies have shown that using a simple parametric technique for signal and foreground subtraction can result in over-subtraction, leading to the loss of the signal.

In this paper, we use ANNs to extract the global 21-cm signal parameters along with foreground and ionospheric parameters from the composite all-sky averaged signal, containing foreground and ionospheric effects. This study considers perfect instrument conditions, representing an ideal scenario in which the instrument is assumed not to modify the signal. In the first case of study, we follow the $\rm\tanh$ parametrized model and Accelerated Reionization Era Simulations (ARES) code \citep{mirocha2012optimized} to construct the cosmological signal. We use the $\rm \log{(T)}-\log{(\nu)}$ polynomial model to map the bright, dominant foregrounds. According to \citet{Harker2016}, a 3rd or 4th-order polynomial is sufficient to map the sky's spectrum. In contrast, \citet{bernardi2015foreground} demonstrated that when adding the antenna's principal chromatic beam, a 7th-order polynomial is required. We followed \citep{datta2016effects} to add the ionospheric effect into the simulated signal and foreground. In this, we consider mainly three effects induced as a resultant: refraction, absorption, and thermal emission, and all these are directly proportional to the electron density (TEC) and temperature of the electrons at various layers of the Ionosphere ($\rm T_{e}$). These ionospheric effects introduce two more parameters into our training data sets. To check and validate the robustness and reliability of the developed model, we have considered a minute variation to the parameters TEC and $\rm T_{e}$ to generate our final training data set. To further check and validate the ANN model's robustness, we use a more realistic set of global 21-cm signals presented in \citep{choudhury2021using} instead of parametrized global 21-cm signals. This global 21-cm signal data has different parameters than the $\rm \tanh$ parametrized global signal. In section \ref{ section 2} of this paper, we briefly review about 21-cm signal. Section \ref{ section 3} mentions the details about the foreground model that we used to map the galactic and extragalactic sources. Section \ref{ section 4} discusses the ionospheric effects and their impacts on the global 21-cm signal observation. We briefly discuss the ANN overview and matrices we used to evaluate the performance of our ANN model in section \ref{ section 5}. Section \ref{section: 6} describes the methodology and procedures to simulate the global 21-cm signal, foreground, and ionospheric effects for training and testing the ANN model. We present the results obtained by our model for all the cases in section \ref{ section 7}. In the last section, we summarize our work and discuss the implications of our predictions.

\section{global 21-cm Signal} 
\label{ section 2}
The 21-cm line of the neutral hydrogen is formed as a result of the hyperfine splitting of the 1S ground state caused by the interchange of the magnetic moments of the proton and electron. The quantity we can measure is known as "differential brightness temperature", $\rm \delta{T_{b}}$. We measure this quantity relative to Cosmic Microwave Background (CMB) followed by \citep[][]{furlanetto2006effects}:
\begin{equation} \label{eq1}
\begin{split}
{\delta{T}_{b} \equiv {T}_{b} - {T}_{\gamma}} 
\end{split}
\end{equation}

\begin{equation} \label{eq2}
\begin{split}
{\delta{T}_{b}} & {= \frac{T_{S}-T_{R}}{1 + z}(1-e^{-\tau_{\nu}})} \\
&  {\approx 27x_{\ion{H}{i}}(1+\delta_{b})\left(\frac{\Omega_{b}h^{2}}{0.023}\right)\left(\frac{0.15}{\Omega_{m,0}h^{2}} \frac{1+z}{10}\right)^{\frac{1}{2}}} \\
& {\left(1-\frac{T_{\gamma}(z)}{T_{s}}\right)\left[\frac{\partial_{r}\nu_{r}}{(1+z) H(z)}\right]^{-1} mK},
\end{split}
\end{equation}

where $\rm x_{\ion{H}{i}}$ is the hydrogen neutral fraction, $\rm \delta_{b}$ represents the fractional over-density in baryons, $\Omega_{m}$ and $\Omega_{b}$ signify total matter density and baryon density, respectively, $\rm H(z)$ is the Hubble parameter, $\rm T_{\gamma}(z)$ denotes CMB temperature at redshift $\rm z$, and $\rm T_{s}$ is spin temperature, and $\partial_{r}\nu_{r}$ is the velocity gradient along the line of sight. 

The 21-cm global signal is a sky averaged signal that offers information on global cosmic occurrences. It can tell us about the story of the thermal history of ionizing radiation like UV radiation which interrupts neutral hydrogen, X-rays that heat the gas and elevate $\rm  T_{k}$, and $\rm  L y_{\alpha}$, which is accountable for the Wouthuysen-Field coupling \citep{wouthuysen}. In the study, the peculiar velocity and density fluctuation components in the global signal (Eqn. \ref{eq2}) are neglected since they average out to a linear order and amount to a minor correction. As a result, the density, neutral fraction, and spin temperature all affect the form of the global signal \citep{2014Mirocha}. 

\begin{equation} \label{eq3}
\begin{split}
{\delta{T}_{b} \approx27(1-x_{i})\left(\frac{\Omega_{b}h^{2}}{0.023}\right)\left(\frac{0.15}{\Omega_{m,0}h^{2}}\frac{1+z}{10}\right)^{\frac{1}{2}}\left(1-\frac{T_{\gamma}(z)}{T_{s}}\right)} 
\end{split}
\end{equation}
 To construct the global signal, we primarily use this equation (\ref{eq3}). 

The spin temperature influences the detectability of the 21-cm signal. Three main quantities determine spin temperature: 
(i) absorption/emission of 21-cm photons by CMB radiation; (ii) collisions with other hydrogen atoms, free electrons, and protons; (iii) scattering of $\rm L_{y}\alpha$ photons that cause a spin-flip through intermediate excitation. In this given limit, the spin temperature sky-average defined as \citep{field1959spin}: 
\begin{equation} \label{eq4}
\begin{split}
{{T_{s} = \frac{T_{\gamma} + y_{\alpha}T_{\alpha} + y_{c}T_{K}} {1 + y_{\alpha} + y_{c}}}},
\end{split}
\end{equation}
where $\rm T_{\gamma}$ is the temperature of radio background, primarily CMB, $\rm T_{\alpha}$ is the color temperature of ambient Lyman-alpha photons, and $\rm T_{K}$ is kinetic gas temperature. $\rm y_{\alpha} $, $\rm y_{k}$ represents the coupling coefficient, which arises due to atomic collision and Lyman-alpha scattering.

\section{Foreground}
\label{ section 3}
The bright foregrounds, mostly caused by galactic and extragalactic sources, are the greatest observational obstacle in observing the global 21-cm signal for studying the CD-EoR. The radio emission from galactic and extragalactic sources is substantially brighter than the global 21-cm signal. We used a very basic model termed log polynomial ($\rm log(T)-log(\nu)$) to map the foreground, which is based on \citet{pritchard201221, bernardi2015foreground}. In our study, we constrain our foreground model to a 3rd order polynomial in $\rm \log(T)-\log(\nu)$, followed by \citet{Harker2016, choudhury2020extracting} which depicts diffuse foregrounds:

\begin{equation} \label{eq5}
\begin{split}
{\ln T_{FG} = \sum_{i=0}^{n=3} a_{i}\left[ \ln (\nu/\nu_{0}\boldsymbol{)}\right]^{i}} ,
\end{split}
\end{equation}

where $\rm a_{0}, a_{1}, a_{2},....,a_{n}$ denote foreground model parameters and $\nu_{0}$ is arbitrary reference frequency. In this study, the derived value of the foreground parameters ($\rm a_{0}, a_{1}, a_{2}, a_{3} $)= ($3.3094$, $-2.42096$, $-0.08062$, $0.02898$) are taken from \citet{harker2015selection, de2008model} and reference frequency taken around $\nu_{0} = 80$ MHz followed by \citep{choudhury2020extracting}. We varied these foreground parameters to construct the different realization of foregrounds (see in the Tab (\ref{tab2})).

\section{Ionospheric Effects}
\label{ section 4}
The ionosphere is a region of the Earth's atmosphere that has a high concentration of electrically charged atoms and molecules. The Sun is one of the most powerful energy sources in the Solar System. Its intense Ultraviolet (UV) and X-ray radiation interact with the Earth's atmosphere to create the ionosphere through photo-ionization. The electron density and temperature change significantly depending on the type of solar fluctuations. Any electromagnetic signal travelling through an optically thin medium, such as the ionosphere, obeys the radiative transfer equation. To understand these effects, the ionosphere is divided into various layers, such as D-layers ($60-90$ km), E-layers, and a composite F-layer ($160-600$ km) \citet{datta2016effects}.

\subsection{F-layer refraction}
The F-layer of the ionosphere, located between $\sim 200$ to $\sim 400$ km above the Earth's surface, accounts for most of the ionospheric electron column density. Outside of this layer, the electron density is very low compared to the inside. Although the electron density varies within the F-layer, we consider it a homogenous shell $200$ to $400$ km in height and assuming a constant electron density of $ \sim \rm 5\times 10^{11}$ $\rm electrons / m^{3}$ which resulting column density of $10$ TEC units \citep{vedantham2014chromatic, datta2016effects}. Due to density differences between the layers, any incoming beam experiences Snell's refraction at the boundaries of the F-layer. The ionosphere's refraction acts like a spherical lens, deflecting incoming light towards the zenith \citep{vedantham2014chromatic, datta2016effects}. As a result of this refraction, any radio antenna located on the ground captures the signal from a wider area of the sky, resulting in a higher antenna temperature.

The angular deviation experienced by any incoming ray with angle $ \rm \theta$ to the horizon in the parabolic layer, which is surrounded by free space with refractive index  $ \rm \eta = 1$, can be calculated as follows \citep{bailey1948new, vedantham2014chromatic, datta2016effects} :

\begin{equation} \label{eq9}
\begin{split}
{\delta\theta(\nu, t)} &{= \frac{2d}{3R_{E}} \left(\frac{\nu_{p}(t)}{\nu}\right)^{2}\left(1+\frac{h_{m}}{R_{E}}\right)}\\
& {\times \left(sin^{2}\theta + \frac{2h_{m}}{R_{E}}\right)^{-3/2}cos\theta},
\end{split}
\end{equation}

where $\rm R_{E} = 6378$ km is the Earth's radius, $\rm h$ represents the altitude, $\rm h_{m}$ represents the altitude where the electron density is maximum in the F-layer, which is $\rm h_{m} = 300$ km, and $\rm d$ represents the change in altitude with respect to $\rm h_{m}$ where the electron density is zero, which is 200 km in our simulation and $\rm \nu_{p}$ is the plasma frequency \citep{thompson2001synthesis}.

\begin{figure*}
\centering
\subfloat[]{\includegraphics[width=7.5 cm]{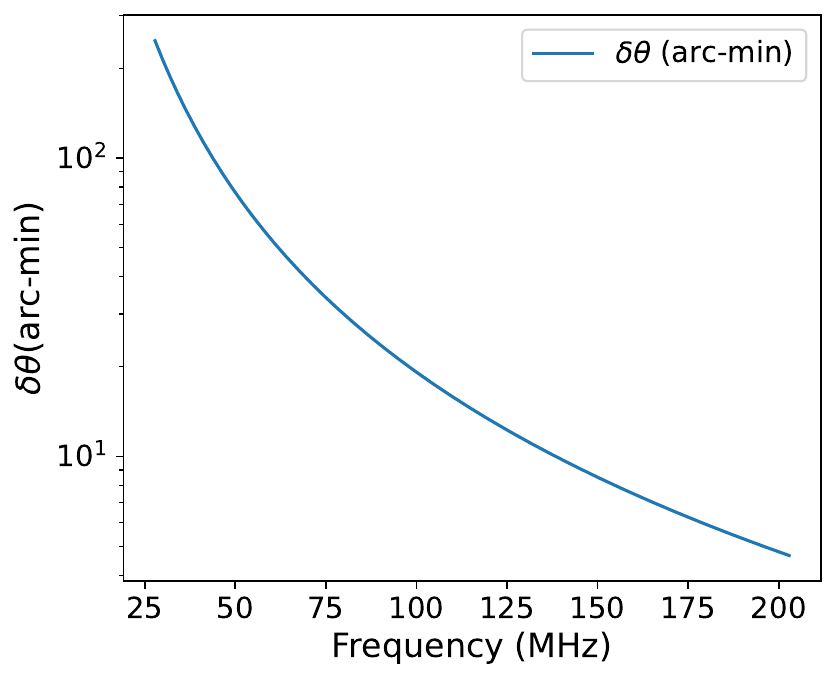}}
\hspace{1.5cm}
\subfloat[]{\includegraphics[width=7.5 cm]{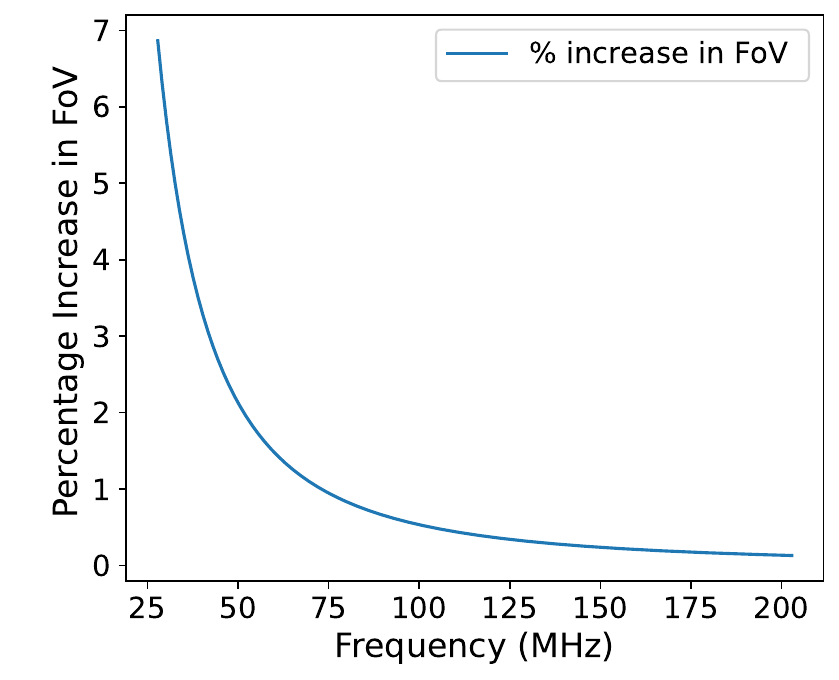}}
\caption{ (a) The deviation angle $\rm \delta\theta$ is plotted as a function of frequency for a typical mid-latitude daytime TEC value (TEC = $10$ TECU). (b) The percentage increase in the field of view  as a function of frequency for the same TEC value.}
\label{fig1}
\end{figure*}

As seen from equation (\ref{eq9}), the ionospheric refraction is proportional to $\rm \nu^{2}$, with the greatest deviation happening for the horizon ray, which has an incidence angle of $0$. As a result of this ionospheric refraction, the field of view at a particular observation frequency will be larger than the primary beam of the antenna. The ionospheric refraction's impact on the angular deviation, as shown in Fig. (\ref{fig1}a) and increase in the field of view (FoV) is calculated and plotted across the frequency ($\rm \nu$), as shown in Fig. (\ref{fig1}b). The resultant antenna temperature, which includes ionospheric refraction, as described by \citet{vedantham2014chromatic}.

\begin{equation} \label{eq10}
\begin{split}
{T_{sky}^{iono}(\nu, TEC(t), \Theta_{0}, \Phi_{0})} & {= \int_{0}^{2\pi} d\Phi} \\
 & {\times \int_{0}^{\pi/2} d\Theta B'(\nu, t;\Theta- \Theta_{0} - \delta\theta(t), \Phi)} \\ 
 & {\times T_{sky}(\nu, t; \Theta_{0}, \Phi_{0}) sin\Theta},
\end{split}
\end{equation}

where $\rm T_{sky}^{iono}$ refers to the antenna temperature that considers ionospheric refraction, ($\rm \Theta_{0}$, $\rm \Phi_{0}$) is the pointing centre. $\rm B'(\nu, t;\Theta- \Theta_{0} - \delta\theta(t), \Phi)$ describes a modified field of view caused by the ionosphere's refractive effect, and $\rm T_{sky}( \nu , \Theta , \Phi)$ denotes actual sky temperature which includes signal and foreground.

\subsection{D-layer absorption and thermal emission }
The D-layer is the lowest layer of the ionosphere, extending from $\sim 60$ to $\sim 90$ km above the Earth's surface \citep{vedantham2014chromatic}. Due to solar insolation, high electron concentrations in the D-layer are projected to last only during the daytime. At night-time, residual electron concentrations are mostly found of the order of $ \sim 10^{8}$\,$\rm electron / m^{3}$.

The high concentration of atmospheric gas in the D-layer at these heights results in significant electron collision frequencies, which cause radio wave attenuation (\citet{evans1968radar}; \citet{davies1990ionospheric}). The absorption by the D-layer can be expressed follows \citep[][]{evans1968radar, datta2016effects}:

\begin{equation} \label{eq13}
\begin{split}
{L_{dB}(\nu, TEC_{D}) = 10 * log_{10} \left(1 + \tau(\nu, TEC_{D})\right)}
\end{split}
\end{equation}

where $\rm TEC_{D}$ signifies the D-electron layer's column density and $ \rm \tau$ indicates the optical depth.

\begin{figure*}
\centering
\subfloat[]{\includegraphics[width=7.5 cm]{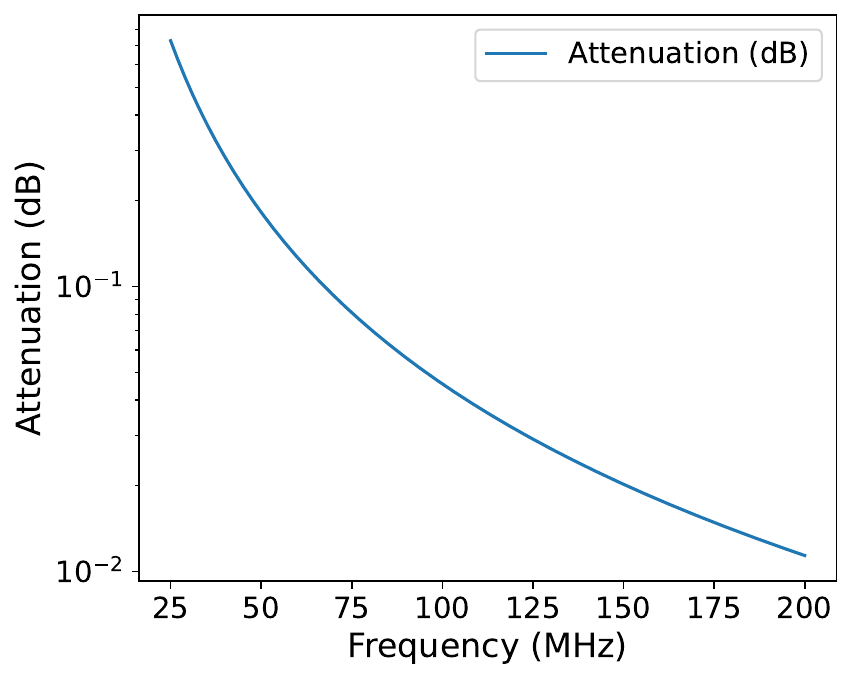}}
\hspace{1.5cm}
\subfloat[]{\includegraphics[width=7.5 cm]{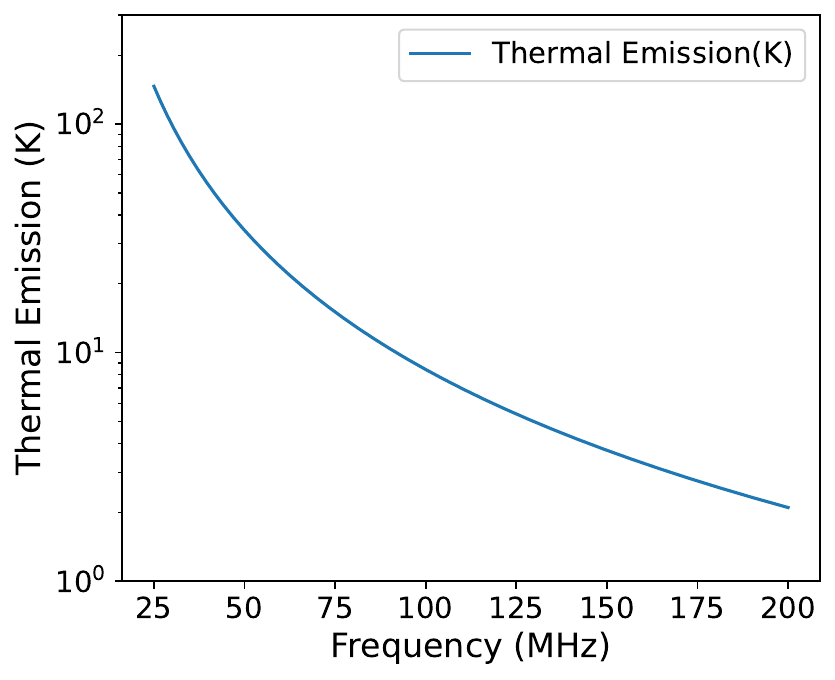}}
\caption{ (a) For the TEC value (TEC= 10 TECU), attenuation is plotted as a function of frequency in the solid line. (b) The variation in thermal emission from the ionosphere is also depicted in the solid line.}
\label{fig2}
\end{figure*}

The D-layer is also responsible for thermal emission \citep[][]{pawsey1951ionospheric,steiger1961observations, hsieh1966characteristics, datta2016effects}, which is included as a $\rm \tau( \nu , TEC(t))<T_{e}>$ into the final term [see in Eq. (\ref{eq6})]. The terms $\tau (\nu, TEC(t))$  represents optical depth for the corresponding ionosphere, and $\rm <Te>$ is average electron temperature, which causes thermal radiation. In our simulations, we consider mid-latitude ionosphere, and we take D-layer electron temperature $ \rm T_{e} = 800 $ K \citep{zhang2004midlatitude}. We have calculated the attenuation factor and thermal emission for the corresponding mid-latitude ionosphere and plotted them against the frequency ($\nu$), shown in Fig.(\ref{fig2}a) and Fig.(\ref{fig2}b). In the plot, we see that as we go lower in frequency ($\nu$), this attenuation factor and thermal emission increase compared to the higher frequency ($\nu$).

Finally, the brightness temperature of the radio signal recorded by the ground-based radio antenna in the presence of all three ionospheric effects is defined as \citep{datta2016effects}:
\begin{equation} \label{eq6}
\begin{split}
{T_{Ant}^{iono}(\nu, TEC(t), \Theta_{0}, \Phi_{0})} & {= T_{sky}^{iono}(\nu, t; \Theta_{0}, \Phi_{0})} \\ 
 & {\times (1 - \tau (\nu, TEC(t))} \\
 & {+\tau (\nu, TEC(t))  * <T_{e}>},
\end{split}
\end{equation}

where $\rm T_{Ant}^{iono}$ is the effective brightness temperature captured by any ground-based antenna, $ \rm T_{sky}^{iono}$ denotes the changed sky brightness temperature as a result of ionospheric refraction, and $(\Theta_{0}, \Phi_{0})$ are pointing center.

\section{Basic Overview of Artificial Neural Network  }
\label{ section 5}

An ANN is a computational algorithm inspired by human biological neural networks. A basic architectural neural network is made up of three primary layers: an input layer, a hidden layer, and an output layer. The number of hidden layers defining its depth and the number of neurons in those layers determines the network width. In a feed-forward network, each neuron in a layer is coupled to every neuron in the next layer, and the information flow is unidirectional. The connections between neurons are associated with weights and biases \citep{ANN2000}.

To describe the detailed structure of the ANN architecture, we followed \citet{shimabukuro2017analysing, choudhury2020extracting}. We considered that there is an $\rm n$ input training data set ($ \rm x_{1},x_{2}.....,x_{n}$). Each input data set is fed by particular neurons in the input layer. For example, the input data $\rm x_{j}$ is provided to the jth neurons in the input layer, which is further connected to the next layer neurons ( hidden layer) with associated a weight $\rm w^{(1)}_{ij}$ and a bias $\rm b^{1}_{j}$. In general, this can be described as:
\begin{equation} \label{eq14}
\begin{split}
{z_{i} = \sum^{n}_{j=1}  w^{\boldsymbol{(}1\boldsymbol{)}}_{ij}x_{j} + b^{1}_{j}}
\end{split}
\end{equation}

In the hidden layer, the output from Eqn. (\ref{eq14}) is further activated by a non-linear activation function $\rm h$, such that $\rm y_{i} = h(z_{i})$. The final output $\rm Y^{'}_{i}$  which is the linear combinations of the activated outputs of the neurons in the hidden layer with weights $\rm w^{(2)}_{ij}$ and biases $\rm b^{2}_{j}$ can be described as \citep{shimabukuro2017analysing}:

\begin{equation} \label{eq15}
\begin{split}
{Y^{'}_{i} = \sum^{n}_{j=1} w^{\boldsymbol{(}2\boldsymbol{)}}_{ij}y_{j} + b^{2}_{j}}
\end{split}
\end{equation}

After each forward pass, a cost or error function is computed at the output layer. This cost function is optimized during training by back-propagating errors iteratively. We can define the total loss (cost) function of the network as follows:
\begin{equation} \label{eq16}
\begin{split}
{E} = & {\frac{1}{N_{t}}\sum^{N_{t}}_{n=1} E_{n}(w,b)}\\
&= {\frac{1}{N_{t}} \sum^{N_{t}}_{n=1}\left[\frac{1}{N}\sum^{N}_{n=1} \left(Y^{'}_{(i,n)}-Y_{(i,n)}\right)^{2}\right]},
\end{split}
\end{equation}
where $ \rm N_{t}$ represents the number of training epochs, $ \rm N$ represents the number of output data elements, $\rm Y^{'}$ denotes prediction by the ANN, and $\rm Y$ denotes the actual output feature. These weights $\rm (w)$ and the biases $\rm (b)$ are updated at the end of each training epoch by using methods called gradient descent in the following manners described below:

\begin{equation} \label{eq17}
\begin{split}
{\Delta w^{l}_{ij} = w^{l}_{0ij}-\eta \frac{\partial E }{\partial w^{l}_{ij}}= w^{l}_{0ij} - \eta \sum^{N_{train}}_{n=1} \frac{\partial E_{n}}{\partial w^{l}_{(ij)}}}\\
{\Delta b^{l}_{ij} = b^{l}_{0ij}-\eta \frac{\partial E }{\partial b^{l}_{ij}}= b^{l}_{0ij} - \eta \sum^{N_{train}}_{n=1} \frac{\partial E_{n}}{\partial b^{l}_{(ij)}}},
\end{split}
\end{equation}

where $w^{l}_{0ij}$ and $b^{l}_{0ij}$ represent the initial weights and bias, respectively, and $\eta$ is the learning rate. We employed Python and the Sequential Model from the Keras API in our feed-forward network. To develop our network, we utilized standard sci-kit learn \citet{pedregosa2011scikit} and Keras modules. We pick the number of hidden layers and number of neurons such that we can get optimum network performance. The number of neurons in the output layer is the same as the number of output parameters we want to predict. The ANNs architecture employed in our study is discussed in detail in the following sections. 

\subsection{ $\rm R^2$ and RMSE Scores}
\label{ section 5.1}
 We choose $\rm R^{2}$ and root mean square error (RMSE) scores as a metric to evaluate network performance. The coefficient of $\rm R^{2}$ and RMSE is obtained for each parameter from the test set of the predictions. The $\rm R^{2}$ scores is defined as:
 
\begin{equation} \label{eq25}
\begin{split}
{R^{2} = \frac{\sum(y_{pred}-\Bar{y}_{orig})^{2}}{\sum(y_{orig}-\Bar{y}_{orig})^{2}} = 1 - \frac{\sum(y_{pred}-y_{orig})^{2}}{\sum(y_{orig}-\Bar{y}_{orig})^{2}}},
\end{split}
\end{equation}
where $\Bar{y}_{orig}$ is the average of the original parameter, the sum is that the score $\rm R^{2}$ = 1 denotes a flawless inference of the parameters across the whole test set, whereas $\rm R^{2}$ might range between 0 and 1.

We have followed \citet{shimabukuro2017analysing} to calculate the normalized RMSE score for prediction :

\begin{equation} \label{eq26}
\begin{split}
{RMSE = \sqrt{\frac{1}{N_{pred}} \sum_{i=1}^{N_{pred}}\left(\frac{y_{orig}- y_{pred}} {y_{orig}}\right)^{2}}} ,
\end{split}
\end{equation}

where $\rm N_{pred}$ represents the total number of samples in prediction data sets, a lower RMSE value suggests that the parameter prediction is more accurate.

\section{Building Of Training and Test Data Sets}
\label{section: 6}
We follow the steps below to construct the data sets for all the different realizations to combine them to build the final training data sets. We created 360 sets of data sets for all the different realizations for both types of signals, parametrized and physical. These data sets are created using each parameter value sampled randomly and uniformly from the given parameter range by the following Tab.(\ref{tab:1}), Tab.(\ref{tab:1b}) and Tab.(\ref{tab2}). 
We further split these constructed data sets into three chunks for training, validation and testing of the model. In the test set, we add additional thermal  noise for the corresponding observational hour by following the radiometer equation details described in the section (\ref{ section 6.4}). 

\subsection{Simulation methods for the global 21-cm signal}
\label{section: 6.1}

\textbf{Case 1: parametrized model}

\begin{figure*}
\centering
\subfloat[Global 21cm Signals.]{\includegraphics[width=8.0 cm]{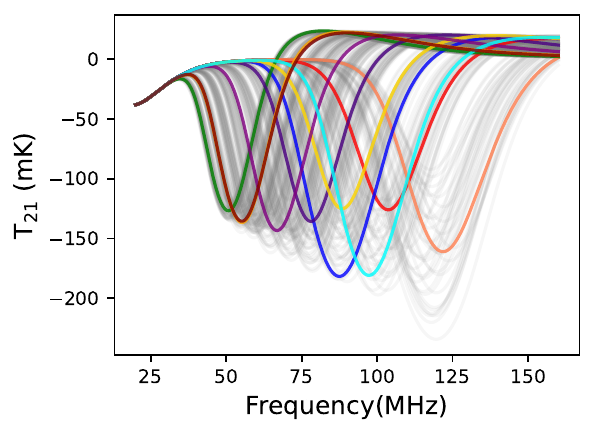}}
\qquad
\subfloat[Global 21cm signals with added foreground.]{\includegraphics[width=8.0 cm]{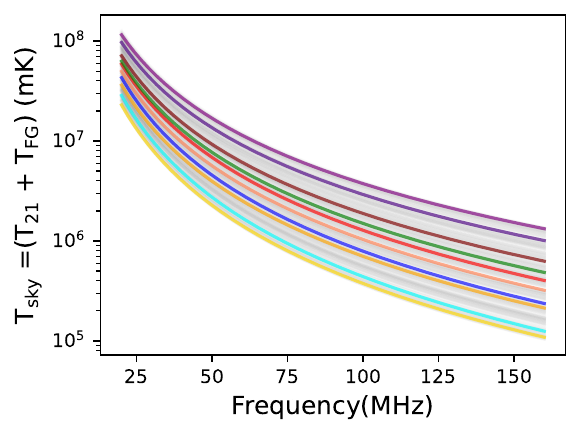}}
\qquad
\subfloat[Global 21cm signals and foreground with ionospheric refraction.]{\includegraphics[width=8.0 cm]{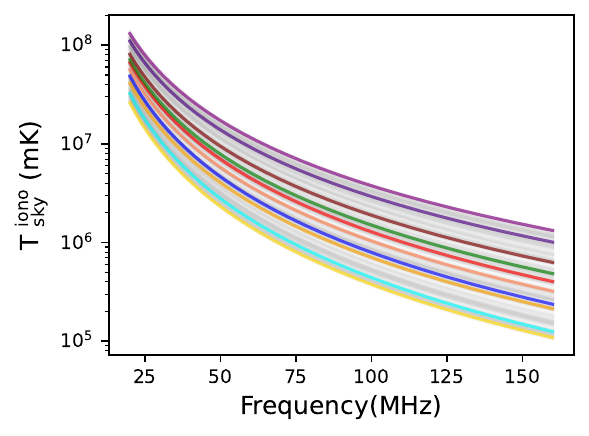}}
\qquad
\subfloat[Excess temperature cause by ionospheric refraction.]{\includegraphics[width=8.0 cm]{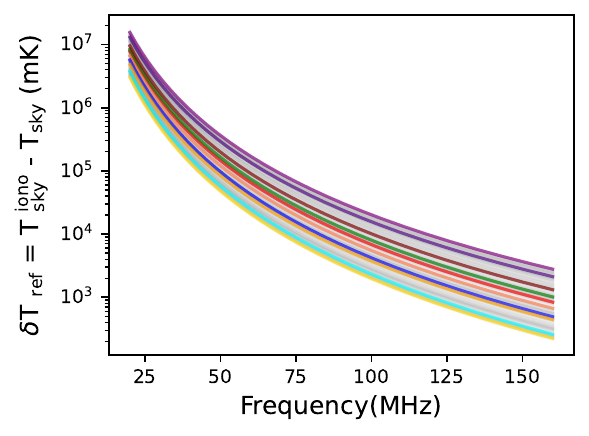}}
\qquad
\subfloat[Global 21cm signals and foreground with all three ionospheric  effects.]{\includegraphics[width=8.0cm]{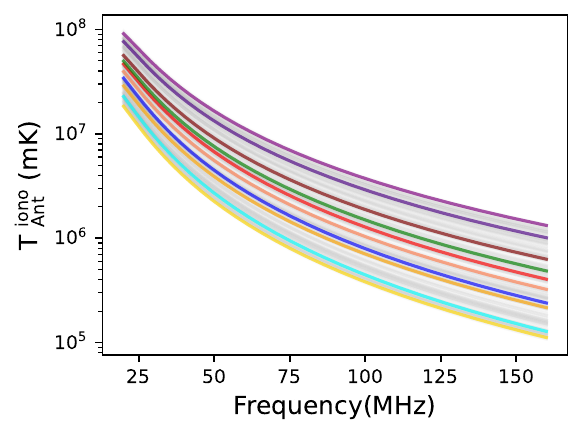}}
\qquad
\subfloat[Contribution of all three ionospheric  effects.]{\includegraphics[width=8.0 cm]{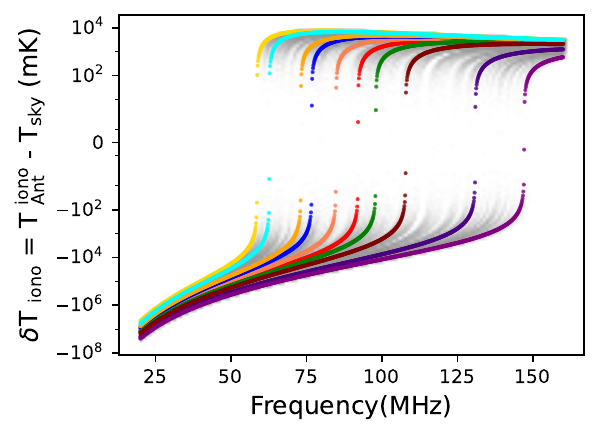}}

\caption{(a) The training data sets for the global 21-cm signals generated using parametrized model by varying the signal parameters.(b) The training data set is constructed by adding foreground into the global 21-cm signals, here clearly we can see how the foreground dominated over the signals. (c) The training data sets for the case when we included the ionospheric refraction effect into the signal and foreground for corresponding fixed TEC value (TEC =10 TECU). (d) The amount of excess temperature recorded by antenna due to ionospheric refraction. (e) The samples of the training data set are constructed by adding all three ionospheric effects- refraction, absorption, and thermal emission into the signal and foreground for variable TEC and $\rm T_{e}$ values. (f) Contribution due to all ionospheric effects into training data sets. In each subplot, a subset of the training data sets is shown in color, while the remaining training data sets are plotted in the background using light gray color.}
\label{fig3}
\end{figure*}

In the first case study, we used the $\rm \tanh$ parameterization model to replicate the global 21-cm signal across the redshift range $6$ < $z$ < $40$ suggested by \citep[][]{mirocha2015interpreting}. This approach utilizes rudimentary $\rm \tanh$ functions to describe the $\rm Ly\alpha$ background, IGM temperature ($\rm T$), and ionization percentage ($\rm \Bar{X})$ , where $\rm Ly\alpha$ background defines the amount of the Wouthuysen-Field coupling \citep[][]{Harker2016}. Each quantity is allowed to grow as a $\rm \tanh$ function \citep[][]{mirocha2015interpreting} by following given Eq.(\ref{eq:18}):

\begin{equation} \label{eq:18}
\begin{split}
{P(z) = \frac{P_{ref}}{2}\left(1 + tanh\frac{(z_{0}-z)}{\Delta z}\right)},
\end{split}
\end{equation}

where $\rm P(z)$ denotes the $\rm \tanh$ model's primary parameter. $\rm P_{ref}$ is step height, $\rm z_{0}$ is pivot redshift, and $\rm \Delta z$ indicates duration. These are free parameters, and their characteristics are directly linked to IGM features but not to source attributes. That is why this model behaves like an intermediate model, which lies between the physical models and phenomenological models like cubic spline \citep{pritchard2010constraining} or Gaussian \citep{bernardi2015foreground, Bernardi2016} models. Now we evolve the model parameters $\rm J_{\alpha}(z)$ ($\rm Ly\alpha$ background), $\rm T(z)$ (IGM temperature), and $\rm \Bar{X_{i}}$ (ionization fraction) as $\rm \tanh$ function by plugging these parameters into Eqn.(\ref{eq:18}), the details are shown below:

\begin{equation} \label{eq19}
\begin{split}
{J_{\alpha}(z) = \frac{J_{ref}}{2}\left(1 + tanh\frac{(J_{z0}-z)}{J_{dz}}\right)} \\
{\Bar{X}_{i}(z) = \frac{X_{ref}}{2}\left(1 + tanh\frac{(X_{z0}-z)}{X_{dz}}\right)}\\
{T(z) = \frac{T_{ref}}{2}\left(1 + tanh\frac{(T_{z0}-z)}{T_{dz}}\right)},
\end{split}
\end{equation}

where $\rm J_{ref}$ represents Ly$\alpha$ flux ( in order of $\rm 10^{-21}$ erg $\rm s^{-1}$ $\rm cm^{-2} Hz^{-1} sr^{-1}$ ), $\rm J_{dz}$ and $\rm J_{z0}$ both represents Ly$\alpha$ background for corresponding redshift interval $\Delta z$ and for the central redshift $\rm z_{0}$ respectively, $\rm T_{dz}$ and $\rm T_{z0}$ are X-ray heating term for the interval $\Delta z$ and for the central redshift $\rm z_{0}$ respectively, and $\rm T_{ref}$ denote step height corresponding $\rm T(z$) parameter, which is fixed at 1000 K. The exact height of the step is not essential because the signal is saturated with low redshifts. $\rm X_{ref}$ represents the step height corresponding to the ionization percentage, and $\Delta z$ and $\rm z_{0}$ are represented by $\rm X_{dz}$ and $\rm X_{z0}$. Finally, we have seven signal parameters along with two fixed parameters ( $\rm X_{ref} =1.0$, $\rm T_{ref}= 1000$ K) to simulate the global 21-cm signal using the $\rm \tanh$ parametrization \citep[][]{choudhury2020extracting}. To generate a simulated 21-cm global signal, we use ARES to determine the coupling coefficient and enter the parameter values into Eq.(\ref{eq2}). We named this simulated signal as a parametrized global 21-cm signal. The derived value of the parameters is taken from
\citep[][]{Harker2016} : $\rm J_{ref} = 11.69$, $\rm J_{dz} = 3.31$, $\rm J_{z0} = 18.54$, $\rm X_{z0} = 8.68$, $\rm X_{dz} = 2.83$, $\rm T_{z0} = 9.77$, $\rm T_{dz}= 2.82$. To produce our training sets, we modified these values by $50\%$ [ see Tab. \ref{tab:1}]. The number of parameters explored is sufficient to cover a wide spectrum of signal morphologies. Figure (\ref{fig3}a) depicts a typical collection of created signals that we will employ. The idea behind the chosen $\rm \tanh$ model was that it can very well mimic the shape of the Global 21-cm signal and is very well tied to the physical characteristics of the IGM. The $\rm \tanh$ parameters are closely related to the IGM characteristics, although they do not provide knowledge about the source properties. As a result, it lies between the phenomenological turning point model and other fully physical theories. 
\\

\begin{table}
\centering

\begin{tabular}{|c|c|}
\hline
\multicolumn{2}{c|}{Case 1 (parametrized) }\\
\hline
Parameters  & Ranges  \\
 &    \\ \hline
Ly$\alpha$ flux ($\rm J_{ref}$) & 5.85-17.54  \\
Ly$\alpha$ background at $\rm z_{0}$ ($\rm J_{z0}$)  & 9.27-27.81  \\
Ionization step at $\rm z_{0}$ ($\rm X_{z0}$) & 4.34-13.02   \\
X-ray heating term at $\rm z_{0}$ ($\rm T_{z0}$) & 4.89-14.65   \\
Ly$\alpha$ background at $\Delta z$ ($\rm J_{dz}$) & 1.65-4.96   \\
X-ray heating at $\Delta z$ ($\rm T_{dz}$) & 1.41-4.23  \\
Ionization step at $\Delta z$  ($\rm X_{dz}$) & 1.42-4.25  \\ \hline

\end{tabular}

\caption{The range of parameters used to build the training data set for the parametrized case of global 21-cm signals.}
\label{tab:1}
\end{table}

\textbf{Case 2: Physical Model}

In the second case study, we used the same data set as a training data set for the signal that was earlier used in \citep{choudhury2021using} to construct the training data set for the global 21-cm signals. They used different physical models based on a semi-numerical algorithm to produce various realizations of the global 21-cm signals across the redshift range 6 < $z$ < 50. The calculation that had been used in the signal construction closely follows \citep {Fc2006} and \citet{pritchard201221}, and the parameters given as input to the model are the following astrophysical parameters:
\begin{itemize}
    \item  ionizing photons escape fraction $\rm f_{esc}$,
    \item X-ray heating efficiency $\rm f_{xh}$,
    \item star formation efficiency $\rm f_{\star}$,
    \item radio background efficiency parameter, $\rm f_{R}$,
    \item number of Lyman-alpha photons produced per baryon in the interested frequency range, $\rm N_{\alpha}$,
\end{itemize}
They vary these astrophysical parameters in the given range, shown in Tab. (\ref{tab:1b}), to construct the training data set for global 21-cm shown in figure(\ref{fig4} a), the detailed calculation and process described in \citep[][]{chatterjee2019ruling, choudhury2021using}. \citep[][]{choudhury2021using} mentioned that the parameters $\rm f_{x}$ and $\rm f_{xh}$ are highly correlated, so in our case study, we combine these two parameters and take them as a single parameter so that we can improve network performance. In their study, they used two different kinds of global signals the first one when there is no excess radio background $\rm f_{R}= 0$ traditional set of signals, and another case when the excess background is present $\rm f_{R}$ non zero, exotic set of signals.
For this study, we have considered the traditional set of the global 21-cm signals for constructing the training data sets.

% Insert the table with R2 Score and RMSE values

\begin{table}
\centering
\setlength{\tabcolsep}{1 pt}
\begin{tabular}{|c|c|}
\hline
\multicolumn{2}{c|}{Case 2 (Physical)} \\
\hline
 Parameters & Ranges \\
       \\ \hline
Normalization factor ($\rm f_{x}$) * X-ray heating efficiency ($\rm f_{xh}$) & 0.0255-7.9800\\
Radio background efficiency ($\rm f_{R}$)     &  0 ( Traditional)     \\
Radio background efficiency ($\rm f_{R}$)    & 2000-1800 ( Exotic)   \\
Star formation efficiency ($\rm f_{star}$)     &  0.0030-0.0099     \\
Ionizing photons escape fraction ($\rm f_{esc}$)      &  0.06-0.19    \\
Number of Lyman-alpha photons ($\rm N_{\alpha}$)   &  9000-800000   \\ \hline

\end{tabular}

\caption{Parameters range that used construct training data set of the global 21-cm signals for the physical case \citep{choudhury2021using}.}
\label{tab:1b}
\end{table}

\subsection{Simulation of foreground}
\begin{figure*}
\centering
\subfloat[Global 21cm signals.]{\includegraphics[width=8cm]{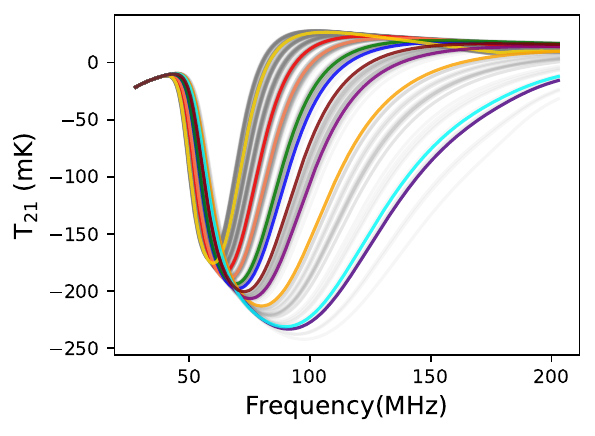}}
\qquad
\subfloat[Foreground added global 21cm signals.]{\includegraphics[width=8cm]{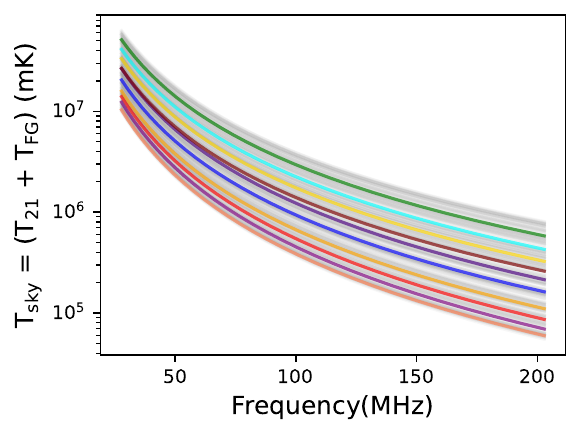}}
\qquad
\subfloat[Global 21cm signals and foreground with ionospheric refraction.]{\includegraphics[width=8cm] {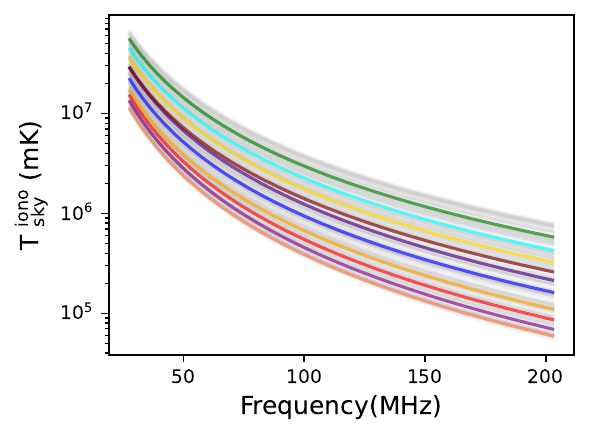}}
\qquad
\subfloat[Excess temperature due to ionospheric refraction.]{\includegraphics[width=8cm]{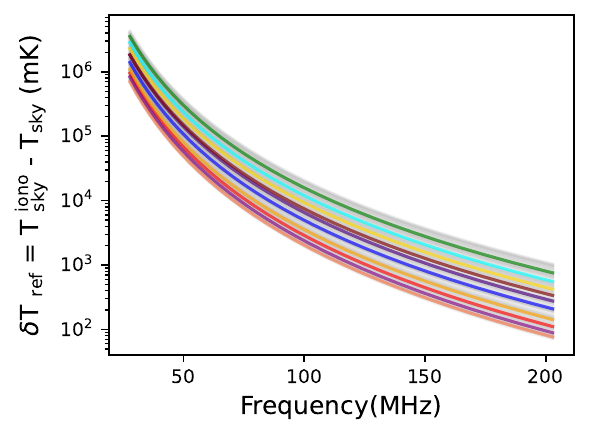}}
\qquad
\subfloat[Global 21cm signals and foreground with all three ionospheric effects.]{\includegraphics[width=8cm]{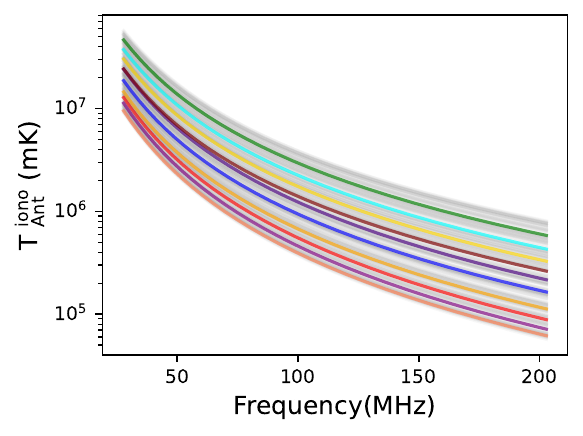}}
\qquad
\subfloat[Contribution due to all the ionospheric effects ]{\includegraphics[width=8cm]{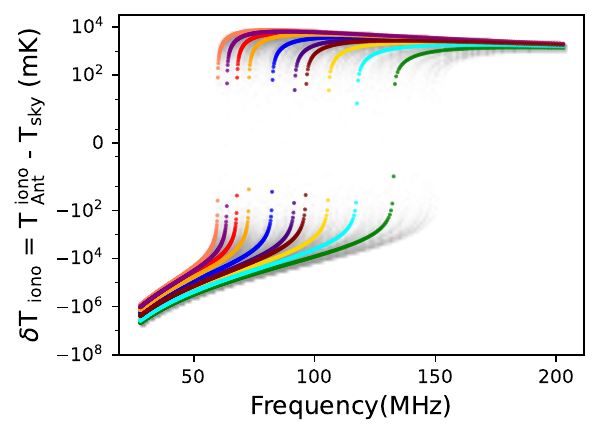}}

\caption{ 
(a) The training data set of the global 21-cm signal was generated using physical model (semi numerical approach). (b)  The training data set after we add foreground into the signal. (c)  The training data set was constructed by including the ionospheric refraction effect into the signal and foreground for the corresponding fixed TEC value 10 TECU. (d) The excess temperature caused by ionospheric refraction as recorded by the antenna in the training data sets.
(e) The samples of the training data set were constructed by adding all three ionospheric effects- refraction, absorption, and thermal emission into the signal and foreground for variable TEC and $\rm T_{e}$ values. (f) Contribution of all ionospheric effects to the training data sets. In each subplot, a subset of the training data sets is shown in color, while the remaining training data sets are plotted in the background using light gray color. }
    \label{fig4}
\end{figure*}

To add foreground into the global 21-cm signal in both cases parametrized and physical, we follow the $ \rm \log(T)-\log(\nu)$ polynomial model described in Section \ref{ section 3}. We simulated foreground by varying its parameters $\rm (a_{0}, a_{1}, a_{2}, a_{3})$ with $( \pm 15\%, \pm 10\%, \pm 1\%, \pm 1\%)$ respectively from its original given value to build the training data for all the scenarios, for more details [see Tab. \ref{tab2}]. Each sample in the training data set (see Fig.\ref{fig3}b) is given by :

\begin{equation} \label{eq:20}
\begin{split}
{T_{sky}(\nu) = T_{21, parametrized}(\nu) + T_{FG}(\nu)},
\end{split}
\end{equation}

where $\rm T_{sky}(\nu)$ is the total sky temperature without including ionospheric effects, $\rm T_{21, parametrized}(\nu)$ is the global 21-cm signal temperature constructed using the parametrized model, $\rm T_{FG}(\nu)$ foreground temperature constructed using the log-log polynomial model.  
For the second case, each sample in the training data set (see Fig.\ref{fig4}b) can be defined as:

\begin{equation} \label{eq:21}
\begin{split}
{T_{sky}(\nu) = T_{21, Physical}(\nu) + T_{FG}(\nu)},
\end{split}
\end{equation}

where $\rm T_{21, Physical}(\nu)$ represents the global 21-cm signal temperature constructed using a semi-numerical physical model.

\subsection{Simulation of ionospheric effects}
To simulate the ionospheric effect, we chose two different scenarios. In the first scenario, we have added only the ionospheric refraction effect for the corresponding fixed TEC value, which is 10 TECU, into the foreground added signal to construct the training data sets shown in Fig.(\ref{fig3}c) and Fig.(\ref{fig4}c). Each sample of the training data set is constructed by the following equation (\ref{eq10}). In the second case study, we have added ionospheric effects, mainly refraction, absorption, and thermal emission, while building the final training data sets shown in Fig.(\ref{fig3}e) and Fig.(\ref{fig4}e). In the final training data set, all the samples are constructed by the following equation (\ref{eq6}).

These ionospheric effects introduce two more parameters in the parameter set: TEC ( Total electron content) and $\rm T_{e}$, representing the thermal temperature of the electron of the D-layer. In our simulation, we have used an F-layer total electron content of TEC 10 TECU and a D-layer electron temperature of $\rm T_{e}$ = 800 K for the mid-latitude ionosphere. We used the International Reference Ionospheric (IRI) model to obtain the TEC value for the D-layer \citep[][]{bilitza2003international}. According to this model, the usual ratio of electron column densities in the D-layer and F-layer is $8.0 \times 10^{-4}$ \citep[][]{datta2016effects}.

For both sets of signals, the parametrized and the physical, to build the final training data sets. We have varied ionospheric parameters by ($\rm TEC$, $\rm T_{e}$) $( \pm1\%, \pm 1\%)$, respectively from their original defined values. The detailed variation that has been used in our simulation to construct the training data set for each case study in this paper is summarized in Tab.\ref{tab2}.

\begin{table}
\centering
\begin{tabular}{lllll}
\hline
Parameters  & Case b & Case c & Case d \\ \hline
Zeroth order foreground coefficient ($\rm a_{0}$)    & $\pm 15\%$        & $\pm 15\%$   & $\pm 15\%$           \\
First order foreground coefficient ($\rm a_{1}$)    & $\pm 10\%$        & $\pm 10\%$   & $\pm 10\%$            \\
Second order foreground coefficient ($\rm a_{2}$)   & $\pm 1\%$         & $\pm 1\%$    & $\pm 1\%$            \\
Third order foreground coefficient ($\rm a_{3}$)    & $\pm 1\%$         & $\pm 1\%$    & $\pm 1\%$             \\
Total electron content ($\rm TEC$)       &                   &   Fixed      & $\pm 1\%$             \\ 
Thermal electron temperature ($\rm T_{e}$)     &                   &      &$\pm 1\%$             \\ \hline
\end{tabular}

\caption{The percentage variation of each parameter of the foreground and ionosphere from its actual value to create upper and lower boundaries and construct the training data set for each scenario for the Case 1 study when we took the parametrized model and case 2 when we consider physical model.}
\label{tab2}
\end{table}

\subsection{Thermal Noise}
\label{ section 6.4}

The thermal noise, $n(\nu)$, in the measured spectrum may be represented as follows using the ideal radiometer equation:
\begin{equation} \label{eq:24}
\begin{split}
{n(\nu) \approx \frac{T_{sys}(\nu)}{\sqrt{\delta \nu  \cdot \tau}}},
\end{split}
\end{equation}
where, $\rm T_{sys}(\nu)$ is system temperature,  $\delta \nu$ is the observational bandwidth  and $\tau$ is the observation time. We are working with simulated observations, which are created using a set of assumptions about signal, foreground, and ionosphere effects. In the future, a similar network will be used to anticipate the redshifted global 21-cm signal based on actual measurements. Actual data from simple observations will replace the with the test data sets.

\section{Results}
\label{ section 7}
In this section, we will discuss results from simulations representing different signal extraction scenarios: signal only, signal with foreground, signal and foreground with ionospheric refraction corresponding to a fixed TEC value, and signal and foreground with all three ionospheric effects with variable TEC and $\rm T_{e}$ values.  We constructed $360$ samples of the data sets for training, validation, and testing of the ANN model for the each following cases. We use $240$ ($67 \%$) samples of the data sets for the training and validation, and the rest of the $120$ ($33 \%$) data sets we used to test the trained ANN model. The validation mainly guides us in tuning the model's hyperparameters, for example, the number of the hidden layer, the number of neurons in the hidden layer, the activation function, the learning rate, etc. It also assists us in identifying overfitting and underfitting by comparing the model loss of the training and validation. In the test set, we add additional thermal noise, $n(\nu)$, corresponding to $1000$ hours of observation by following the radiometer equation (\ref{eq:24}) to construct the final test data set for the each cases of studies.

\subsection{Case 1a: Signal only (parametrized model)}
\begin{figure}
\centering
 \includegraphics[width=8.0 cm]{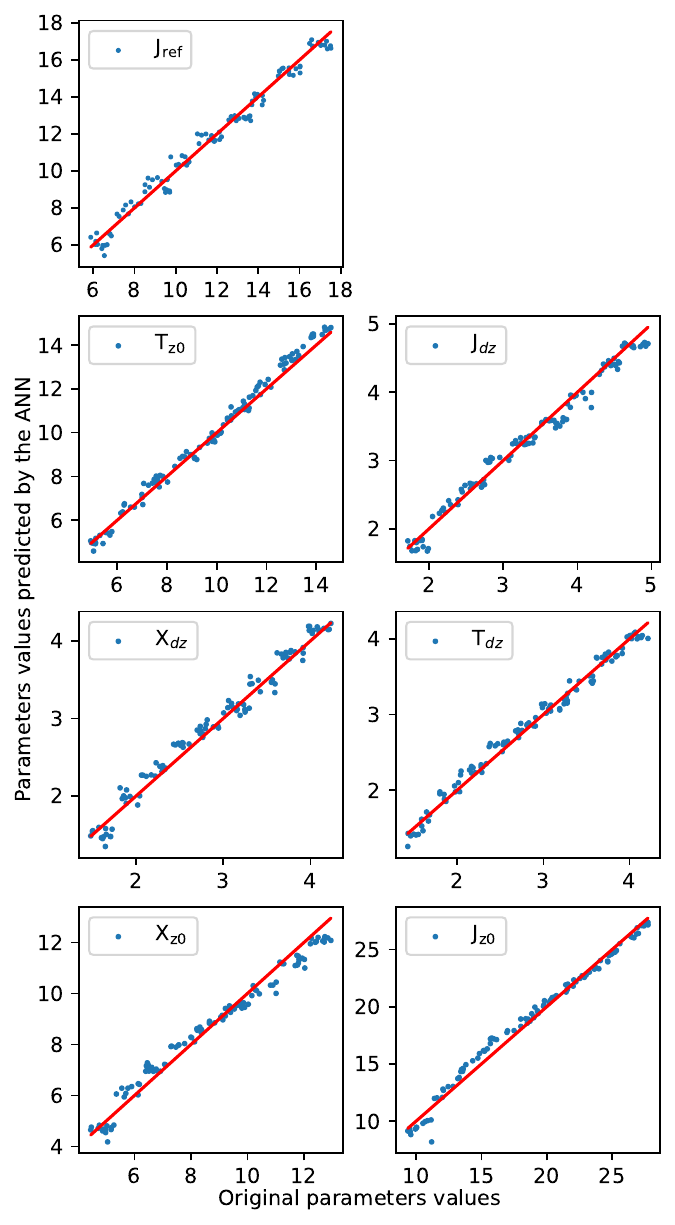}
 \caption{Case 1a: Parmeterized global 21-cm signals. The original values of the parameters are shown by the solid straight line in each plot, while the dots indicate the predicted values by ANN.} 

 \label{fig5}
\end{figure}

\begin{figure}
\centering
 \includegraphics[width=7.5 cm]{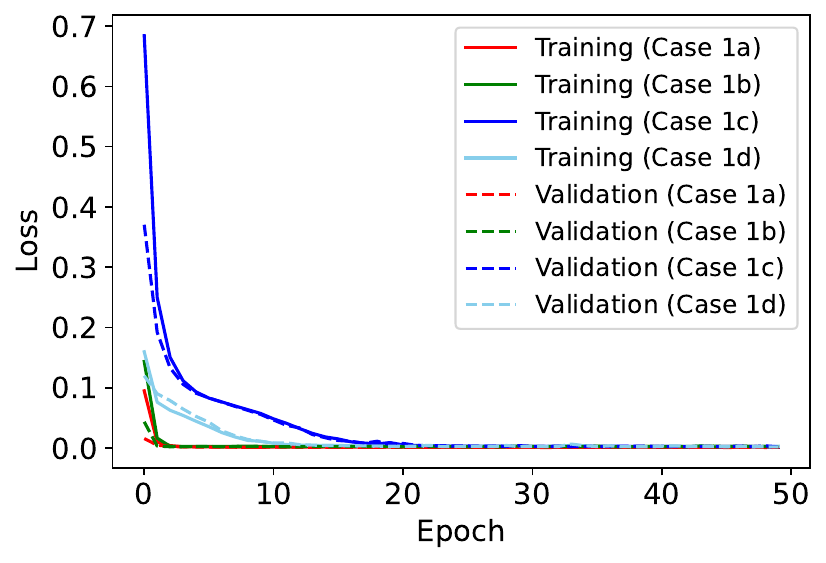}
 \caption{This graph depicts the evolution of the network's loss function when parametrized signals were incorporated. For all situations, the training loss is represented as a solid line, and the validation loss is plotted as a dashed line as a function of epochs. We can see that the test loss function closely follows the training loss function.}
 \label{fig10}
\end{figure}

In the first case, to train our model, we choose the parametrized global 21-cm signals as training data sets, shown in figure (\ref{fig3}a). The model we use for training is constructed with Keras' Sequential API and comprises $1024$ input neurons matching with $1024$ frequency channels and two hidden layers with $16$ and $11$ neurons, respectively, each activated by the 'elu' activation function. The output layer has $7$ neurons to predict the global 21-cm signal parameters. The input training data sets are normalized using the' StandardScaler' function, and corresponding parameters are normalized using 'MinMaxScaler' available in sklearn. We tested our saved model with a test data set and calculated the $\rm R^2$-score and RMSE score for each parameter from the predictions of test sets to figure out how well the network predicts the parameters. The result of $\rm R^2$-score is listed in Tab. (\ref{tab3}) and RMSE scores are listed in Tab. (\ref{tab4}), and the plots of the original versus predicted values of the parameters for the test data set are shown in Fig. (\ref{fig5}). The $\rm R^2$ scores for the predicted signal parameters range from $0.98$ to $0.99$, which shows the network predicted signal parameter is very accurate. To check the overfitting for all the cases, we have plotted training loss and validation loss as a function of the number of epochs [see Fig. \ref{fig10}]. In Fig. (\ref{fig10}), we can see that training loss and validation loss closely flow, and both got converse after 20 epochs for all the cases.

\subsection{Case 1b: Signal with foreground}
\begin{figure*}
\centering
\includegraphics[width=17.0 cm]{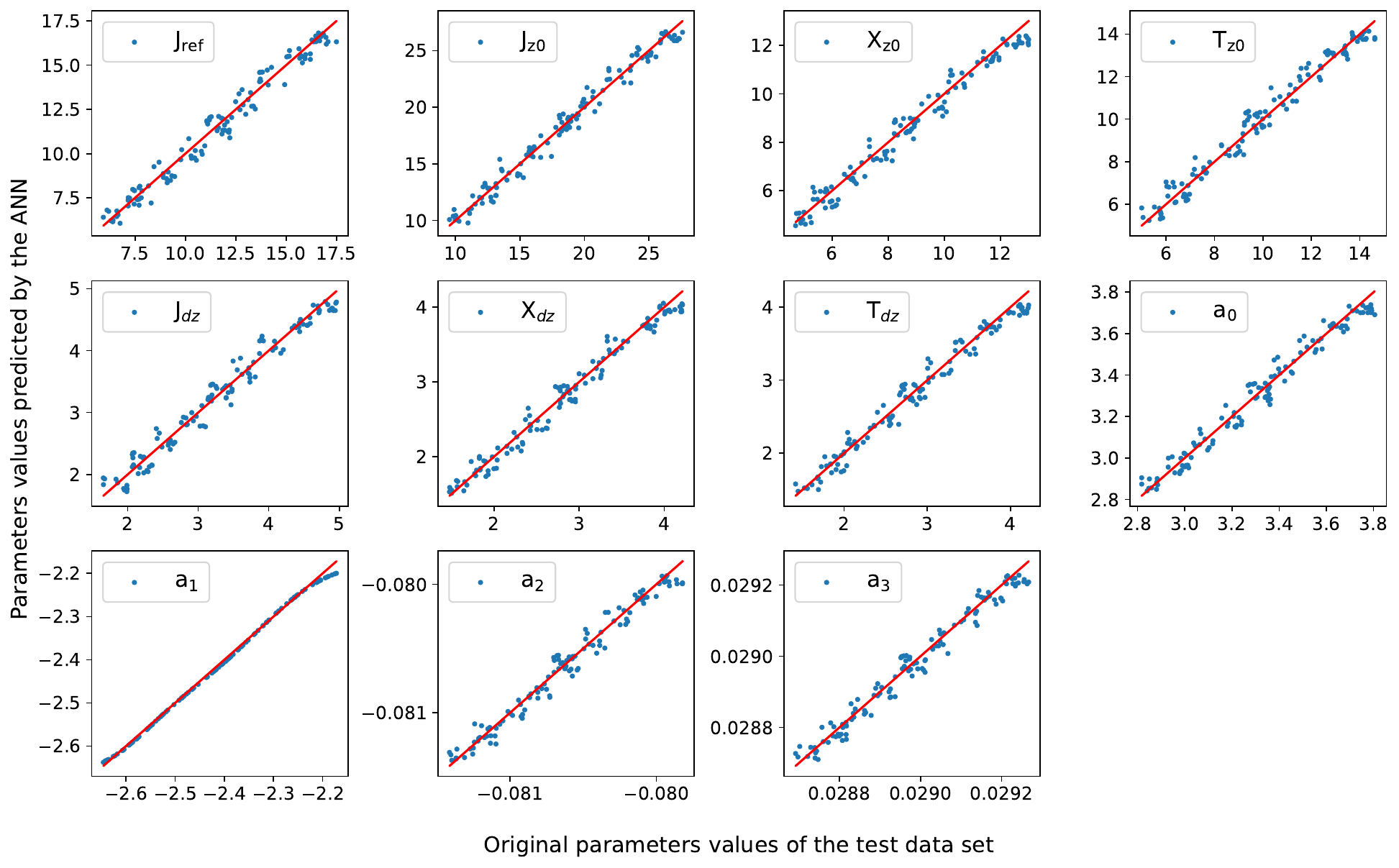}
 \caption{Case1b: parametrized global 21-cm signal with foreground. The original values of the parameters are shown by the solid straight line in each plot, while the dots indicate the predicted values by ANN.}
 \label{fig7}
\end{figure*}

Foregrounds would dominate during the observations of the 21-cm signal for all ground-based experiments. In this case, we train our ANN model with training data constructed by adding foreground to the parametrized global 21-cm signals, shown in Fig. (\ref{fig3}b). In \citep{choudhury2020extracting} has already shown a similar implementation. It is a proof-of-concept to see how effectively our model extracts parameters when adding the foreground. The model architecture we used is different from the first case. The model we have used has $4$ layers made using sequential API from Keras. The input layer has $1024$ neurons that correspond to the $1024$ frequency channel, while the hidden layers have $32$ and $16$ neurons that are activated by the 'sigmoid' activation function. The output layer has $11$ output neurons to predict the global 21-cm signal and foreground parameter. We will use the same model architecture, optimizer, and normalization method for the other cases. The only difference is the number of neurons in the output layer, depending on the number of output parameters.  We calculate the $\rm R^2$-scores and RMSE for each parameter from the test set predictions to determine how well the network predicts the parameters. The plots of the original versus the predicted values of the parameters for the test data set are shown in Fig. (\ref{fig7}), and $\rm R^2$ score and RMSE score for corresponding parameters are listed in Tab. \ref{tab3} and Tab. \ref{tab4}. The $\rm R^2$ score for this case ranges from $0.97$ to $0.98$ for predicted signal parameters, which are significantly lower than the previous case (Case 1a). In the predicted foreground parameters, $\rm a_{0}$ has the highest $\rm R^2$ score of $0.99$. 

% Insert the plots 

% R2 Score for ARES
\begin{table}
\centering
\begin{tabular}{lllll}
\hline
Parameters & Case 1a & Case 1b  & Case 1c  & Case 1d   \\ \hline
$\rm J_{ref}$   & 0.9809  &  0.9778 &  0.9553   & 0.9614        \\
$\rm J_{z0 }$   & 0.9957  &  0.9780  &  0.9442  & 0.9738          \\
$\rm X_{z0}$    & 0.9865  &  0.9760  &  0.9534  & 0.9713          \\
$\rm T_{z0}$    & 0.9911  &  0.9746  &  0.9534  & 0.9668          \\
$\rm J_{dz}$    & 0.9826  &  0.9735  &  0.9482  & 0.9634          \\
$\rm X_{dz}$    & 0.9830  &  0.9848  &  0.9487  & 0.9578          \\
$\rm T_{dz}$    & 0.9874  &  0.9781  &  0.9574  & 0.9595          \\
$\rm a_{0}$     &         &  0.9936  &  0.9757  & 0.9810          \\
$\rm a_{1}$     &         &  0.9738  &  0.9494  & 0.9655           \\
$\rm a_{2}$     &         &  0.9773  &  0.9534  & 0.9586           \\
$\rm a_{3}$     &         &  0.9774  &  0.9568  & 0.9610           \\
$\rm TEC$       &         &          &          & 0.9658            \\ 
$\rm T_{e}$     &         &          &          & 0.9728           \\ \hline
\end{tabular}

\caption{The computed $\rm R^2$-scores for all signal, foreground, and ionosphere parameters for each case studied are listed here. We used the parametrized model to construct the global 21-cm signal.}
\label{tab3}
\end{table}

% RMSE for ARES
\begin{table}
\centering
\begin{tabular}{lllll}
\hline
Parameters & Case 1a & Case 1b  & Case 1c  & Case 1d \\ \hline
$\rm J_{ref}$   & 0.0337  & 0.0440  & 0.0608  & 0.0628        \\
$\rm J_{z0 }$   & 0.0175  & 0.0438  &0.0675   &0.0605          \\
$\rm X_{z0}$    & 0.0329  & 0.0453  &0.0621   &0.0616          \\
$\rm T_{z0}$    & 0.0253  & 0.0470  &0.0645   &0.0577          \\
$\rm J_{dz}$    & 0.0353  & 0.0482  &0.0650   &0.0608          \\
$\rm X_{dz}$    & 0.0370  & 0.0467  &0.0584   &0.0658           \\
$\rm T_{dz}$    & 0.0312  & 0.0434  &0.0642   &0.0636           \\
$\rm a_{0}$     &         & 0.0232  &0.0449   &0.0439          \\
$\rm a_{1}$     &         & 0.0477  &0.0639   &0.0590           \\
$\rm a_{2}$     &         & 0.0443  &0.0621   &0.0642            \\
$\rm a_{3}$     &         & 0.0445  &0.0599   &0.0625            \\
$\rm TEC$       &         &          &        &0.0578            \\ 
$\rm T_{e}$     &         &          &        &0.0588           \\ \hline
\end{tabular}

\caption{The calculated RMSE values for all the signal, foreground, and ionospheric parameters are listed here for each case studied.}
\label{tab4}
\end{table}

\subsection{Case 1c: Signal and foreground with ionospheric refraction}
\begin{figure*}
\centering
\includegraphics[width=17.0 cm]{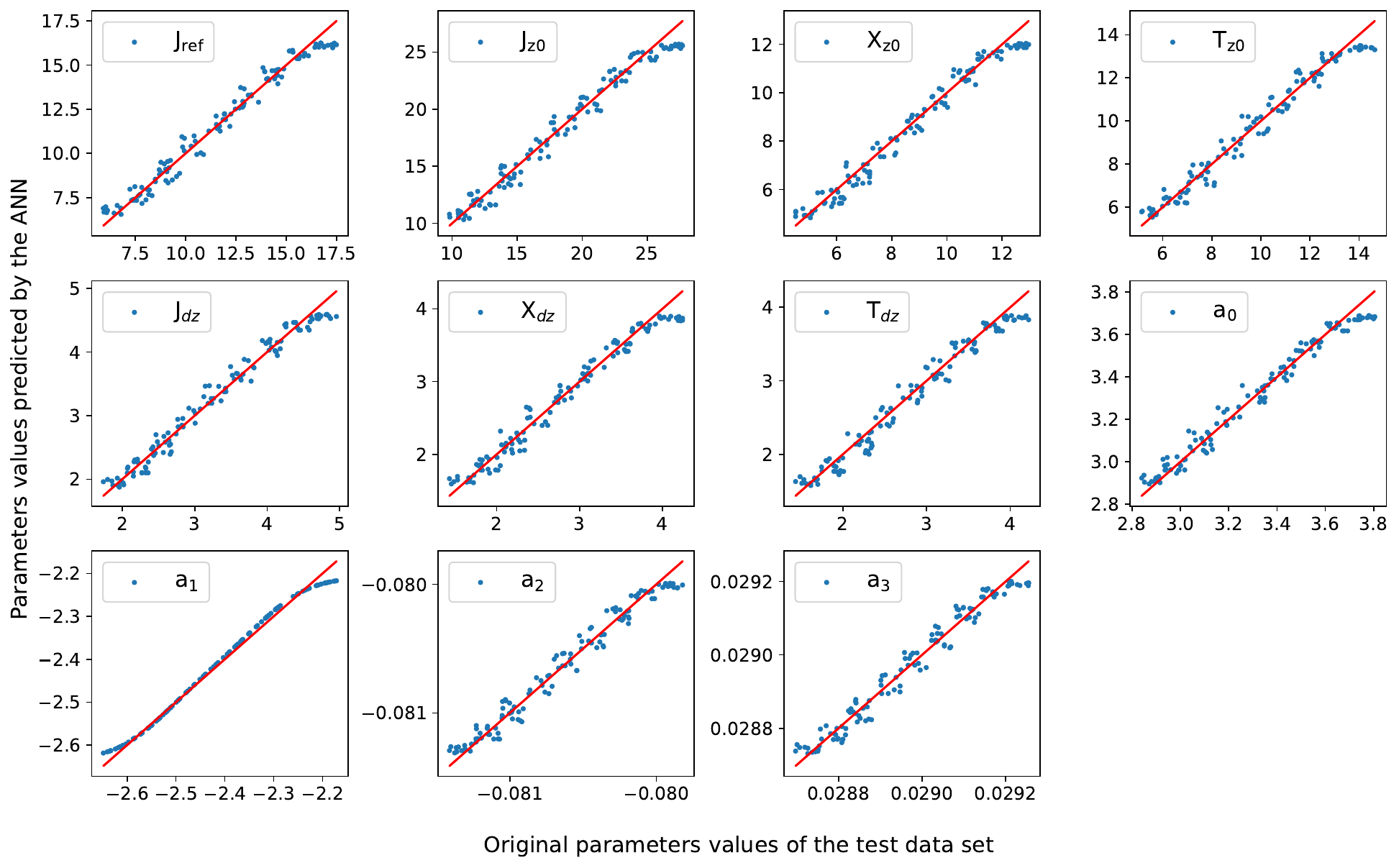}
\caption{Case 1c: parametrized global 21cm signal and foreground with ionospheric refraction for fixed TEC value. The original values of the parameters are shown by the solid straight line in each plot, while the dots indicate the predicted values by ANN.}
 \label{fig8}
\end{figure*}

The ionosphere of the Earth severely distorts low-frequency radio measurements in ground-based observations. We add the ionospheric refraction effects into the signal and foreground while constructing data sets to check this effect, shown in Fig. (\ref{fig3}c). We followed the same ANN model structure that was used in Case 1b and trained the ANN model using training data sets constructed by adding the effects of ionospheric refraction. We tested the trained model with test data sets and calculated the $\rm R^{2}$ score and RMSE score corresponding parameters to evaluate our model performance; the detailed result is listed in Tab. (\ref{tab3}) and Tab. (\ref{tab4}). The predicted values of the parameters by the network are plotted in Fig. (\ref{fig8}). In this case, the obtained $\rm R^2$ score for the signal parameters ranges from $0.94$ to $0.96$.

% Insert the plots 

\subsection{Case 1d: Signal and foreground with all three ionospheric effects- refraction, absorption and thermal Emission}
\begin{figure*}
\centering
 \includegraphics[width=17.0 cm]{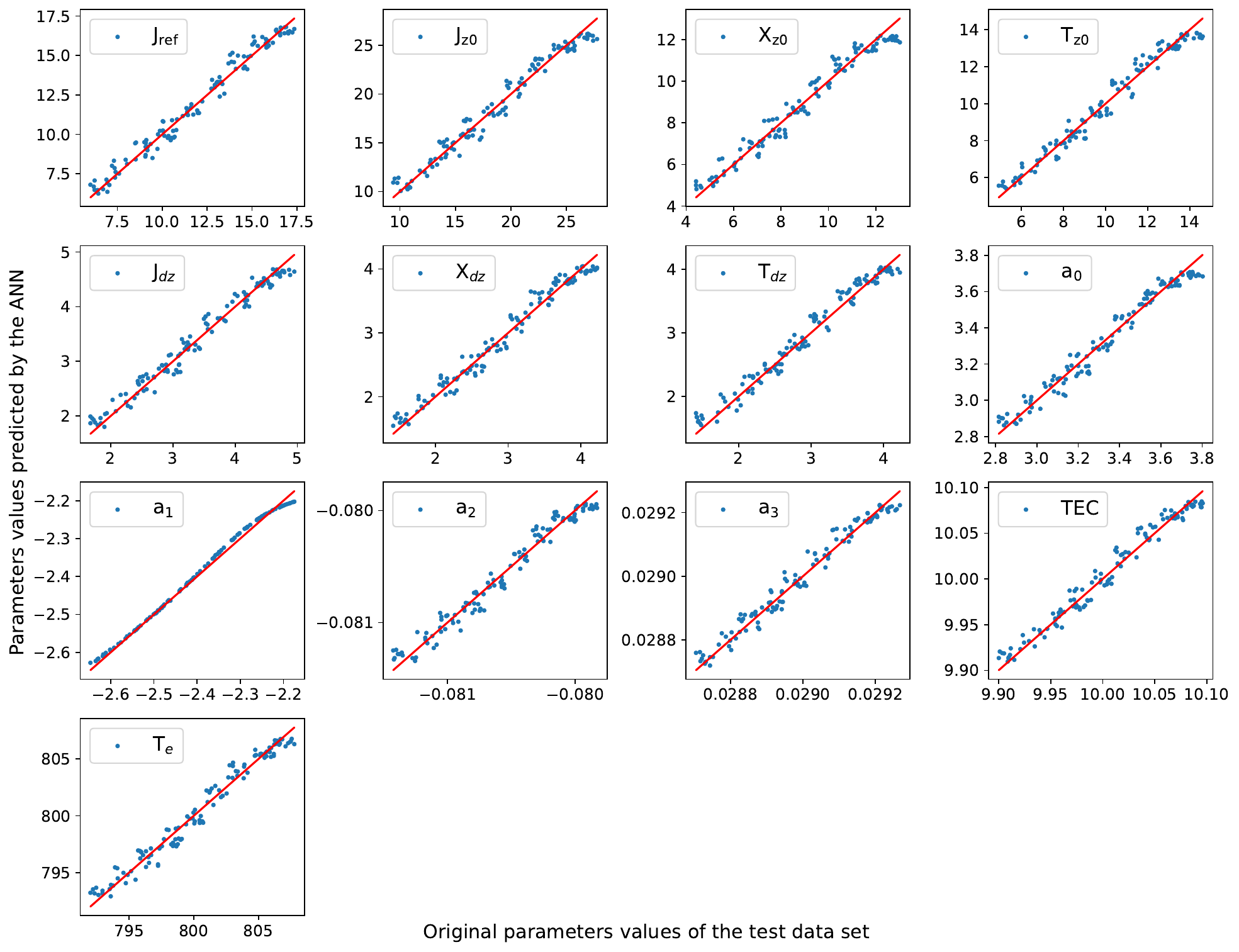}
 \caption{Case 1d: parametrized global 21-cm signal and foreground with all three ionospheric effects- refraction, absorption, and thermal emission. The original values of the parameters are shown by the solid straight line in each plot, while the dots indicate the predicted values by ANN.}
 \label{fig9}
\end{figure*}

In this case, we add other ionospheric effects like absorption and thermal emission and construct the training data set, which we name the "final training data set", shown in Fig. (\ref{fig3}e). We utilized the same architecture as in previous cases (Cases 1b and 1c) to build an ANN model; the only difference here is that the output layer of the model has 13 output neurons corresponding to the 7 signal parameters, 4 foreground parameters, and 2 ionospheric parameters. We test this model with test data and calculate the $\rm R^{2}$ score and RMSE score corresponding parameters to evaluate our model performance. The $\rm R^{2}$ and RMSE scores for each parameter are listed in Tab. \ref{tab3}  and Tab. \ref{tab4}, and predicted values of the parameters by the network are plotted in Fig. (\ref{fig9}). From Table \ref{tab3}, and \ref{tab4}, we can see that $\rm R^2$ values slightly decrease, and the RMSE value slightly increases when we introduce foreground and ionospheric effects into the signals compared to Case 1a when we take signal only. When we add more complexity to the training data set, we have to train our network sufficiently well to maintain high accuracy levels.

% Insert the plots 

\begin{figure}
\centering
 \includegraphics[width=8.0 cm]{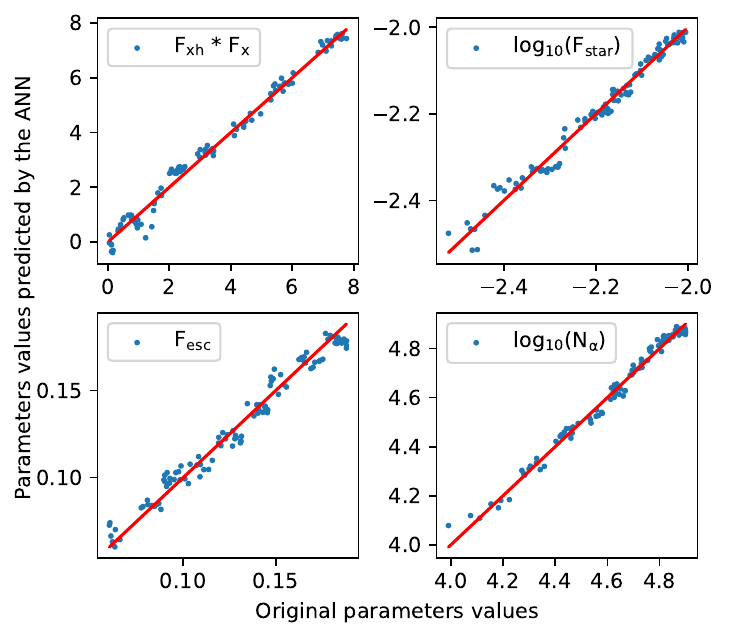}
 \caption{Case 2a: Global 21-cm signal constructed using physical model. The original values of the parameters are shown by the solid straight line in each plot, while the dots indicate the predicted values by ANN. However, $\rm F_{star}$ and $\rm N_{\alpha}$ are plotted in logarithmic scale.}
 \label{fig6}
\end{figure}

\begin{figure}
\centering
 \includegraphics[width=7.5 cm]{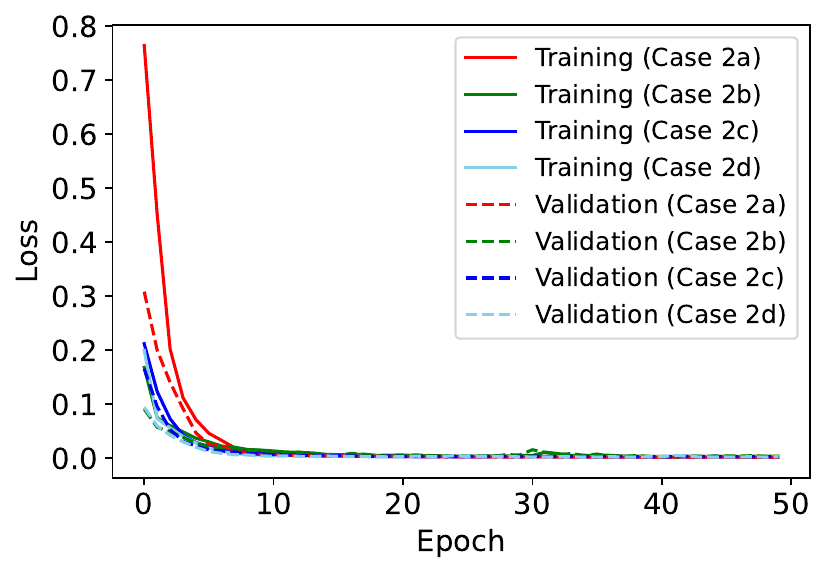}
 \caption{This graph shows the evolution of the network's loss function when we used a signal generated by a physical model. In all cases, the training loss is depicted as a solid line as a function of epochs, whereas the validation loss is plotted as a dashed line.
The test loss function closely matches the training loss function in this case. }
 \label{fig11}
\end{figure}

\subsection{Case 2a : Signal only (physical model)}

To check the robustness and reliability of the network, we trained the ANN model for all the scenarios that we have studied previously with an entirely new set of the global 21-cm signal, taken from \citep{choudhury2021using}. They constructed the global 21-cm signals by using semi-numerical code followed by \citet{chatterjee2019ruling}. It contains a more realistic and diverse group of the 21-cm signal than the parametrized model signals. The signal parameters are also different than the parametrized model; here we use astrophysical parameters ({$\rm f_{xh}*f_{x}$, $\rm f_{star }$,$\rm f_{esc}$, $\rm N_{\alpha}$}). In the first case of the study, we take global 21-cm signals as a training, validation, and testing of the ANN model, shown in Fig. (\ref{fig4}a). The training model consists of a four-layer structure built with Keras' sequential API, with $1024$ input neurons matching $1024$ frequency channels and two hidden layers with $12$ and $11$ neurons activated by the 'elu' activation function.
The output layer contains four output neurons that predict astrophysical parameters of the global 21-cm signal ({$\rm f_{xh}*f_{x}$, $\rm f_{star }$, $\rm f_{esc}$, $\rm N_{\alpha}$}). We used the StandardScaler function, which is available in 'sklearn' to preprocess input signals. At the same time, we use MinMaxScaler to scale the signal parameters before passing them to the model. We use 'adam' as the optimizer and' mean squared error' as the loss function. Once the network is trained and validated, we save the model. We used the same optimizer and normalization method for the other cases.

 We test the model using a test data set and obtain the $\rm R^2$ scores and RMSE for each parameter from the test set predictions to see how well the network predicts the parameters. The predicted parameters are plotted in Fig. (\ref{fig6}) and the corresponding $\rm R^2$ score and RMSE score for each parameter are listed in Tab.\ref{tab5} and Tab. \ref{tab6}. From the Tab.\ref{tab5}, we can see that the parameter $\rm N_{\alpha}$ has the highest $\rm R^2$ scores of $0.99$ and parameters $\rm f_{esc}$ has the lowest $\rm R^2$ score of $0.98$. To check the overfitting for all the cases, we have plotted training loss and validation loss, similar to the parametrized case [see Fig. (\ref{fig11})]. In Fig. (\ref{fig11}), we can see that training loss and validation loss closely flow, and both got converse after 20 epochs for all the cases.

\subsection{Case 2b : Signal with foreground} 
\begin{figure*}
\centering
\includegraphics[width=17.0cm]{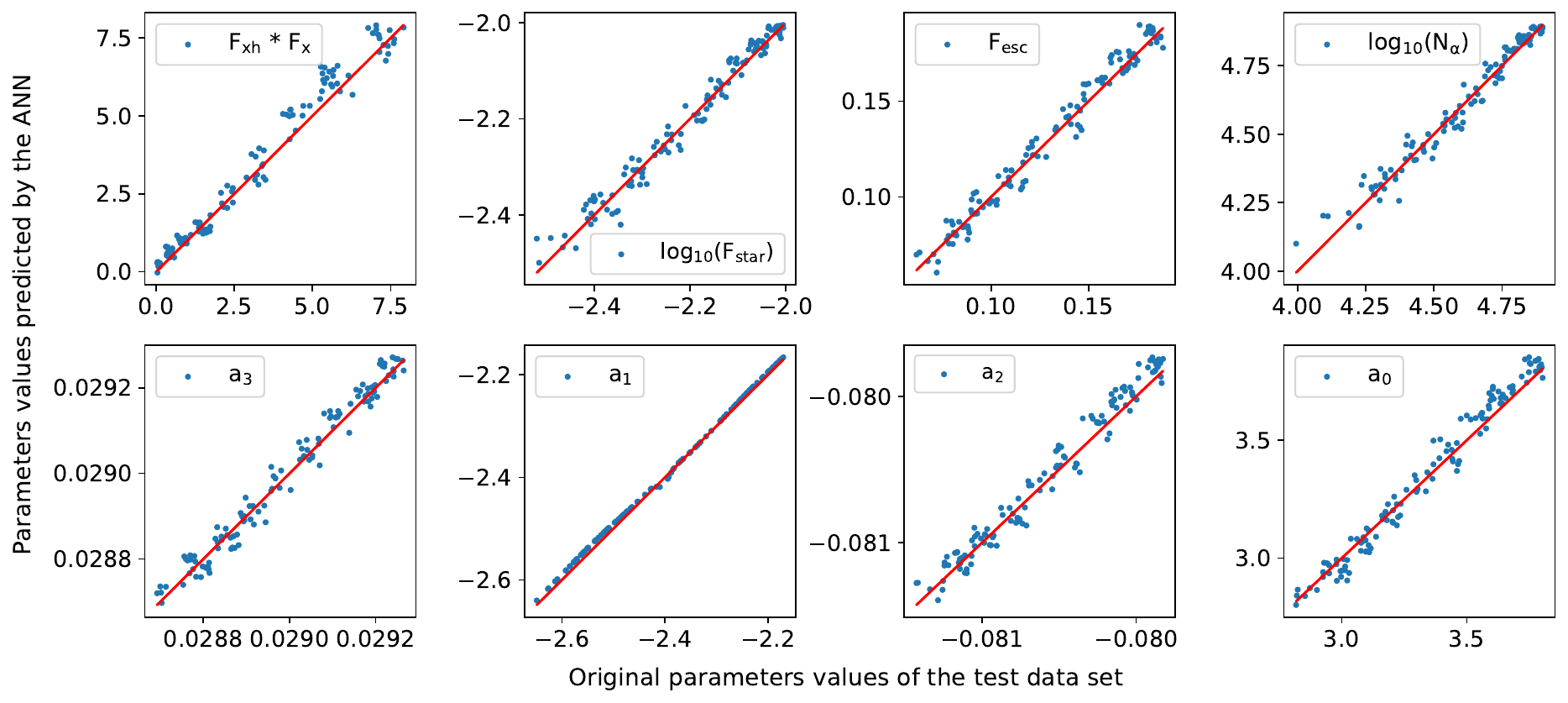}
\caption{Case 2b: Signals with foreground. In the each plots the original values of the parameters are shown by the solid straight line, while the dots indicate the predicted values by ANN. However, $\rm F_{star}$ and $\rm N_{\alpha}$ are plotted in logarithmic scale.}
\label{fig12}
\end{figure*}

In this case, we train our model with a training data set, which is constructed by the combination of the foreground with different realizations of the global 21-cm signal, shown in Fig. (\ref{fig4}b). We employed a $5$-layer model architecture for training, with $1024$ input neurons matching the $1024$ frequency channels and three hidden layers with $32$, $16$ and $16$ neurons activated by the 'elu' activation function. The output layer of the model has $8$ output neurons corresponding to $4$ astrophysical parameters of the global 21-cm signal and $4$ foreground parameters. We test our trained model with the test data set and calculate the $\rm R^2$ and RMSE scores for the corresponding parameters shown in Tab. \ref{tab5} and Tab. \ref{tab6}, and plot of predicted values of the parameters against the original values is shown in Fig. (\ref{fig12}). From Tab. \ref{tab5}, we can see that the foreground parameters' $\rm R^2$ score is much higher than the signal parameters; this means the network predicts the foreground parameter more accurately than the signal parameters. In the signal, the parameter $\rm f_{x,h}* f_{x}$ has the highest $\rm R^2$ scores of $0.98$, and $\rm N_{\alpha}$ has the lowest $\rm R^2$ scores of $0.96$.   

% Insert the plots 
follow

\subsection{Case 2c : Signal and foreground with ionospheric refraction}
\begin{figure*}
\centering
 \includegraphics[width=17.0 cm]{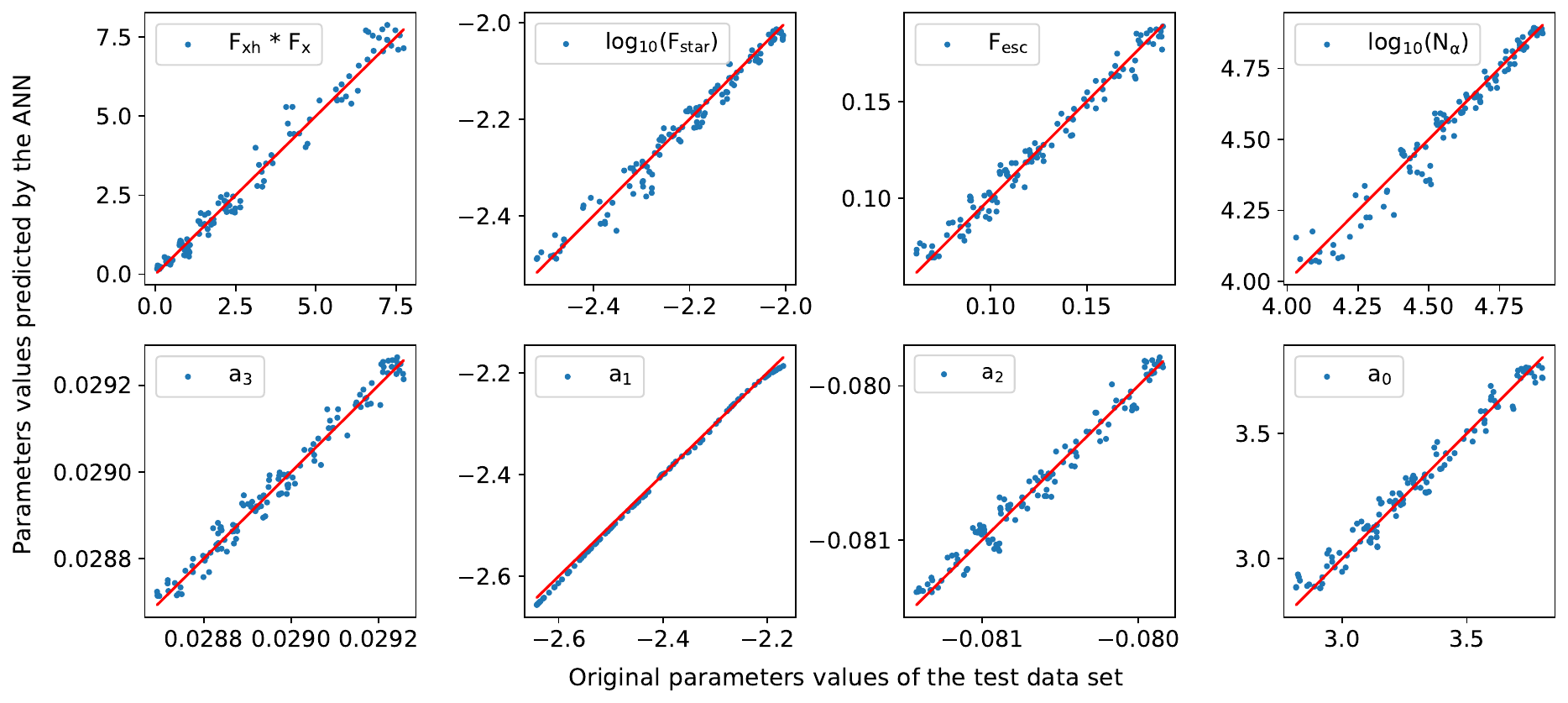}
 \caption{Case 2c: Signal and foreground with ionospheric refraction for fixed TEC value. In the each plots, the original values of the parameters are shown by the solid straight line, while the dots indicate the predicted values by ANN. However, $\rm F_{star}$ and $\rm N_{\alpha}$ are plotted in logarithmic scale.}
 \label{fig13}
\end{figure*}

In the third case, we added foreground and ionospheric refraction effects for corresponding fixed TEC value into the global 21-cm signal to build the training data set, shown in Fig. (\ref{fig4}c), and randomly divide these samples into the same ratio as in Case 2(b) to train and test the ANN model. We follow the same model architecture that we used previously for the signal in the foreground case to build the model for this case. Now we train this model and save it for further validation and testing. We test the saved model with test data sets, and we calculate the $\rm R^{2}$ score and RMSE score for the corresponding parameter. Details of the result are listed in Tab. \ref{tab5} and Tab. \ref{tab6}, and a plot of predicted values of the parameters against the original values is shown in Fig. (\ref{fig13}). The $\rm R^2$ score for the foreground parameters has been improved from the previous case, but the signal parameter $R^2$ score decreases in comparison; this means adding more complexity to the training data making signal extraction more difficult. We obtained the $\rm R^2$ score for the signal parameters around 0.96 to $0.98$ [see Tab. \ref{tab5}].

\subsection{Case 2d : Signal and foreground with all three ionospheric effects- refraction, absorption, and thermal emission}
\begin{figure*}
\centering
 \includegraphics[width=17.2 cm]{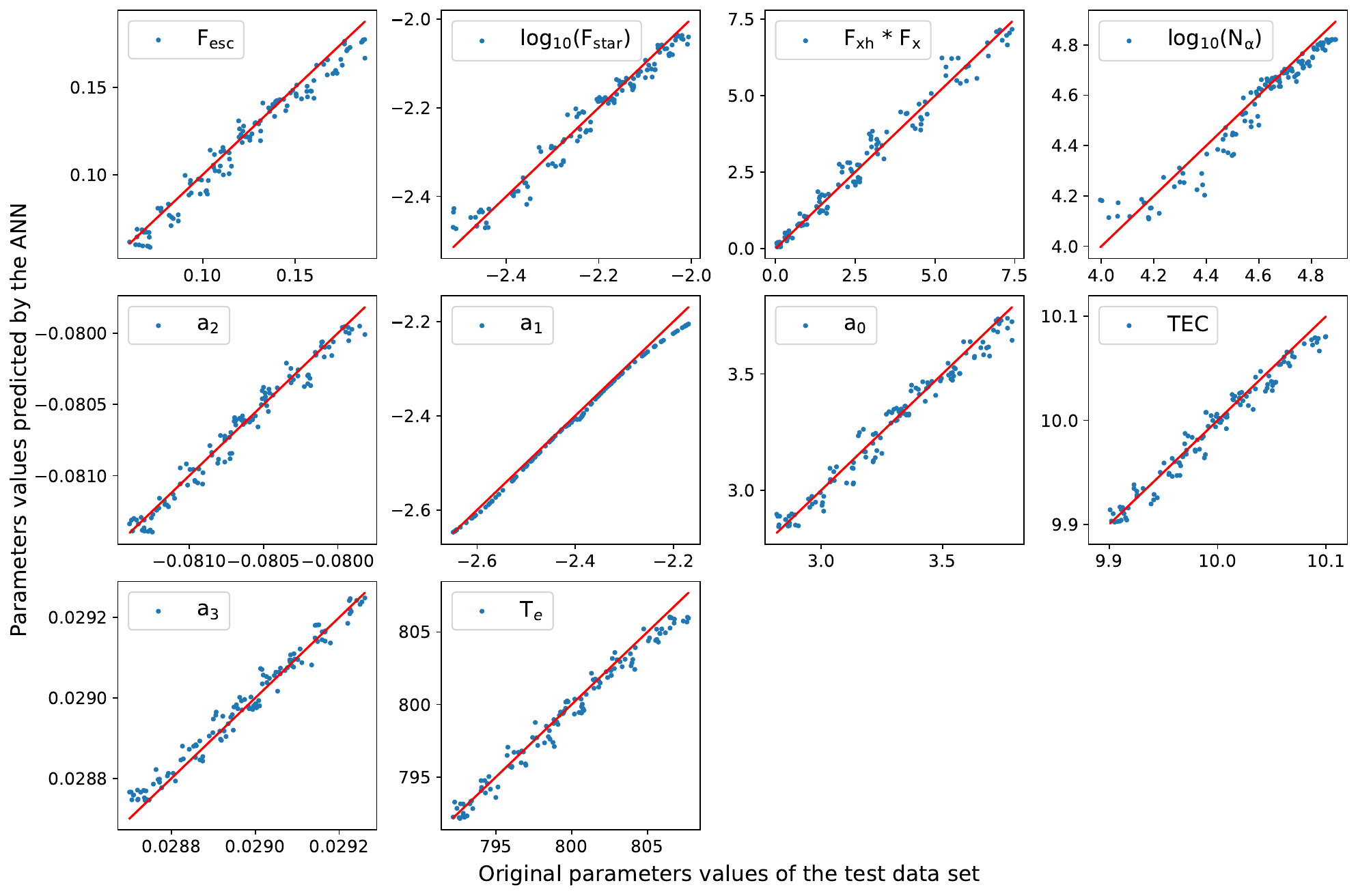}
 \caption{Case 2d: Signal and foreground with all three ionospheric effects-Refraction, Absorption, and Thermal Emission. In the each plots the original values of the parameters are shown by the solid straight line, while the dots indicate the predicted values by ANN. However, $\rm F_{star}$ and $\rm N_{\alpha}$ are plotted in logarithmic scale.}
 \label{fig14}
\end{figure*}

In the last case, we have constructed training data sets by the combination of the global 21-cm signal, foreground, and ionospheric effects (refraction, absorption, and thermal emission), shown in Fig. (\ref{fig4}e). To build the model for this case, we use the same architecture that we used in previous models except for the outer layer. Here, the output layer of the model has $10$ output neurons corresponding to the $4$ signal parameters, $4$ foregrounds, and $2$ ionospheric parameters. Once the models were trained and validated, we saved them for further testing. We tested the saved train model with the test data set and calculated the $\rm R^2$ score and RMSE score for the corresponding parameters to check the network performance. The values of $\rm R^{2}$ score and RMSE score of each parameter are tabulated in Tab. (\ref{tab5}) and Tab. (\ref{tab6}). The plots of the actual versus predicted values of the parameters for the test data set are shown in Fig. (\ref{fig14}). The $\rm R^2$ score for the foreground and ionospheric parameters is much higher than the signal parameters $\rm R^2$ score. The reason is simple: foreground and ionospheric effects dominated the final training data sets signal that we are giving to the network as fed compared to the signals. The $\rm R^2$ score we obtained for the signal parameters is around $0.96$ to $0.98$ [see Tab. \ref{tab5}].
% Insert the plots 

\begin{table}
\centering
\begin{tabular}{lllll}
\hline
Parameters         & Case 2a & Case 2b & Case 2c & Case 2d  \\ \hline
$\rm f_{xh}*f_{x}$ &  0.9906  & 0.9813  & 0.9839     & 0.9804      \\
$\rm f_{star }$    &  0.9893  & 0.9731  & 0.9675     & 0.9704       \\
$\rm f_{esc}$      &  0.9863  & 0.9772  & 0.9756     & 0.9751        \\
$\rm N_{\alpha}$   &  0.9952  & 0.9607  & 0.9579     & 0.9620         \\
$\rm a_{0}$        &          & 0.9752  & 0.9812     & 0.9981          \\
$\rm a_{1}$        &          & 0.9964  & 0.9957     & 0.9990           \\
$\rm a_{2}$        &          & 0.9739  & 0.9773     & 0.9793            \\
$\rm a_{3}$        &          & 0.9736  & 0.9743     & 0.9773             \\
$\rm TEC$          &          &         &            & 0.9733              \\
$\rm T_{e}$        &          &         &            & 0.9742               \\ \hline

\end{tabular}
\caption{The computed $\rm R^2$-scores for all signal, foreground, and ionosphere parameters for each case studied are listed here. We used physical model (semi numerical model) to construct the global 21-cm signal.}
\label{tab5}
\end{table}

\begin{table}
\centering
\begin{tabular}{lllll}
\hline
Parameters      & Case 2a     & Case 2b & Case2c         & Case 2d \\  \hline
$\rm f_{xh}*f_{x}$  & 0.0174 & 0.0428  & 0.0395  &0.0411  \\
$\rm f_{star }$     &0.0287  &0.0476   &0.0540   &0.0488   \\
$\rm f_{esc}$       &0.0334  &0.0453   &0.0507   &0.0475    \\
$\rm N_{\alpha}$    &0.0176  &0.0562   &0.0567   &0.0508     \\
$\rm a_{0}$         &        &0.0473   &0.0410   &0.0442      \\
$\rm a_{1}$         &        &0.0107   &0.0276   &0.0094      \\
$\rm a_{2}$         &        &0.0433   &0.0503   &0.0420      \\
$\rm a_{3}$         &        &0.0481   &0.0533   &0.0445      \\
$\rm TEC$           &        &         &         &0.0480      \\
$\rm T_{e}$         &        &         &         & 0.0473      \\ \hline

\end{tabular}
\caption{The calculated RMSE values for all the signal, foreground, and ionospheric parameters are listed here,for each case studied.}
\label{tab6}
\end{table}

\subsection{ Time varying ionospheric effects- refraction, absorption and thermal emission}

\begin{figure}
\centering
 \includegraphics[width=7.5 cm]{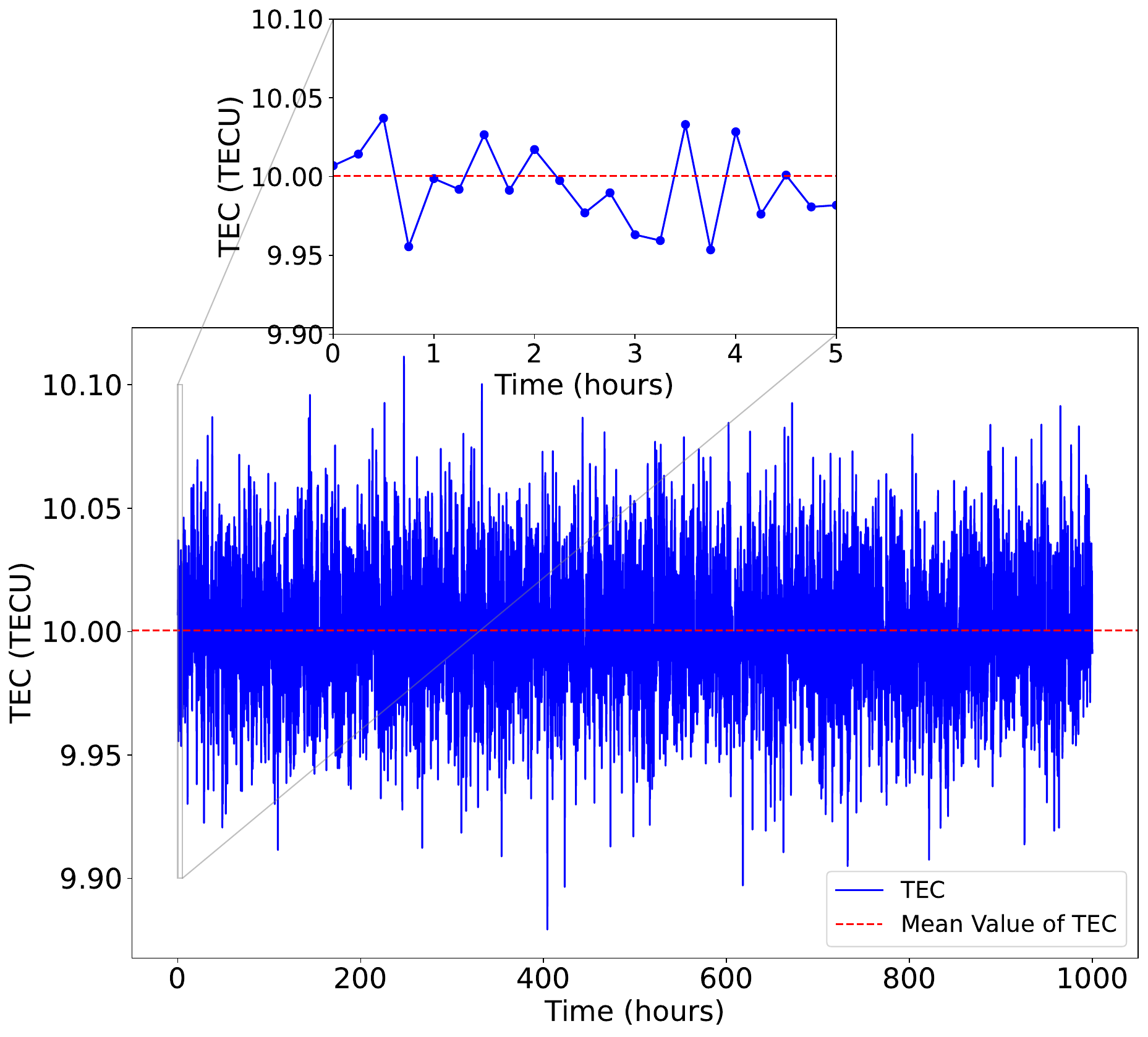}
 \caption{ The blue lines in this graph depict the F-layer Total Electron Content (TEC) variation across a 1000-hour observation period. The red dashed line represents the calculated mean TEC value derived from these fluctuations.}
 \label{fig15}
\end{figure}

\begin{figure}
\centering
 \includegraphics[width=7.5 cm]{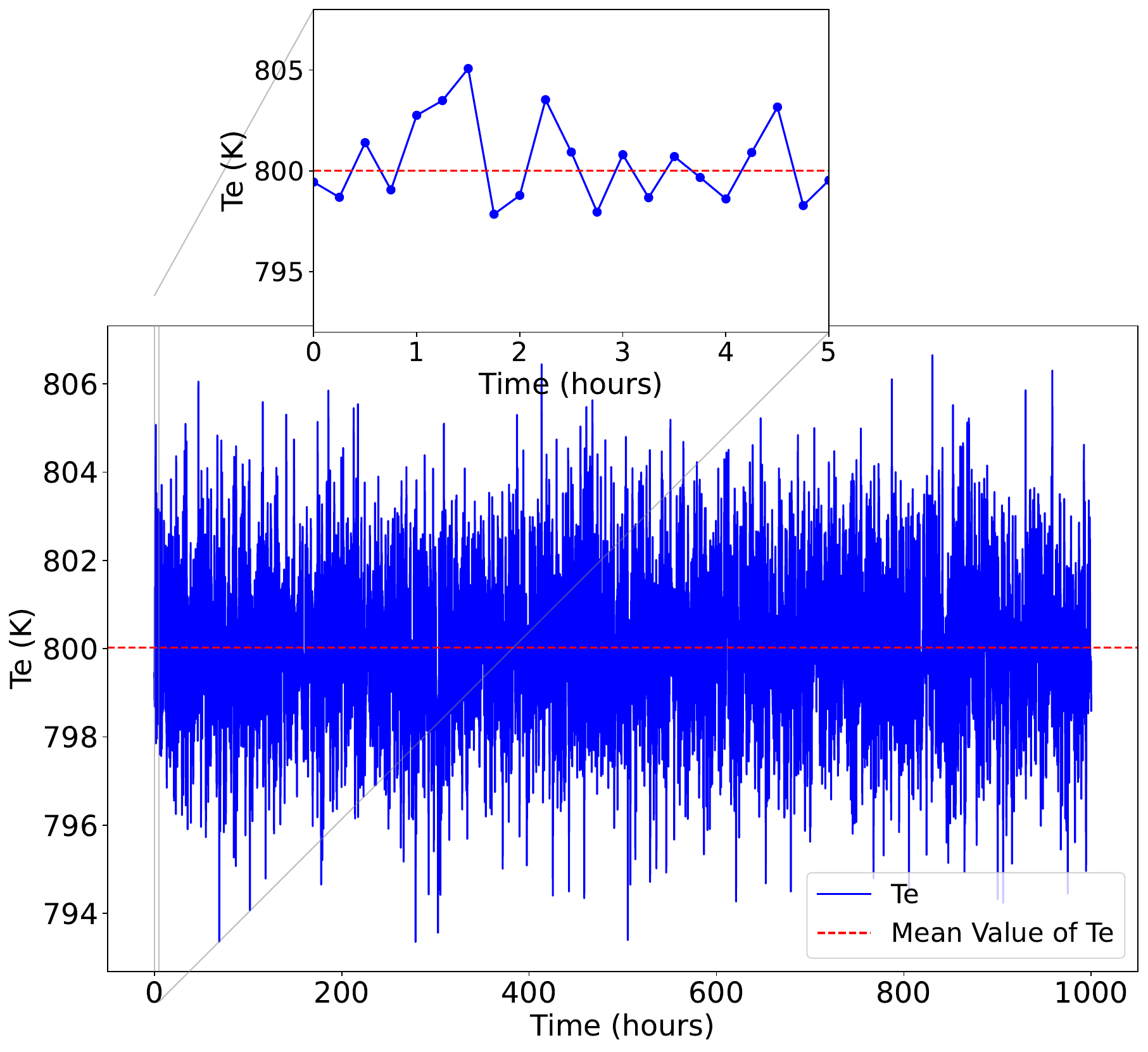}
 \caption{The blue lines in this graph depict the variation of the D-layer electron temperature (Te) across a 1000-hour observation period. The red dashed line represents the calculated mean Te value derived from these fluctuations.}
 \label{fig16}
\end{figure}

\begin{figure}
\centering
 \includegraphics[width=7.2 cm]{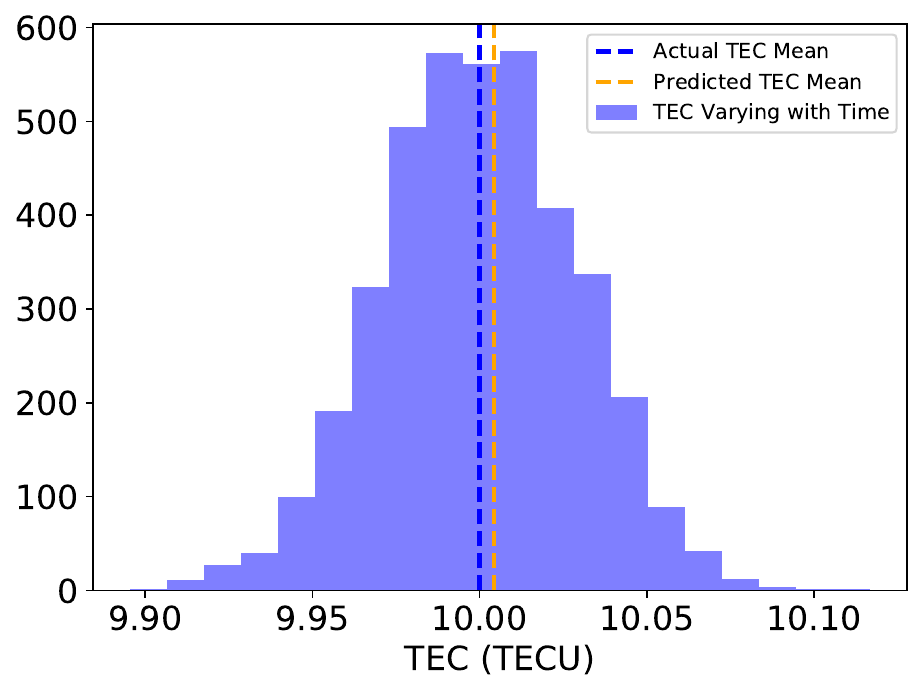}
 \caption{This histogram presents the distribution of TEC values containing the entire observation duration in the context of the parametrized signal scenario. The blue dashed line denotes the average value of the actual TEC, while the orange dashed line corresponds to the mean of the predicted TEC values by the ANN model.}
 \label{fig17}
\end{figure}

\begin{figure}
\centering
 \includegraphics[width=7.2 cm]{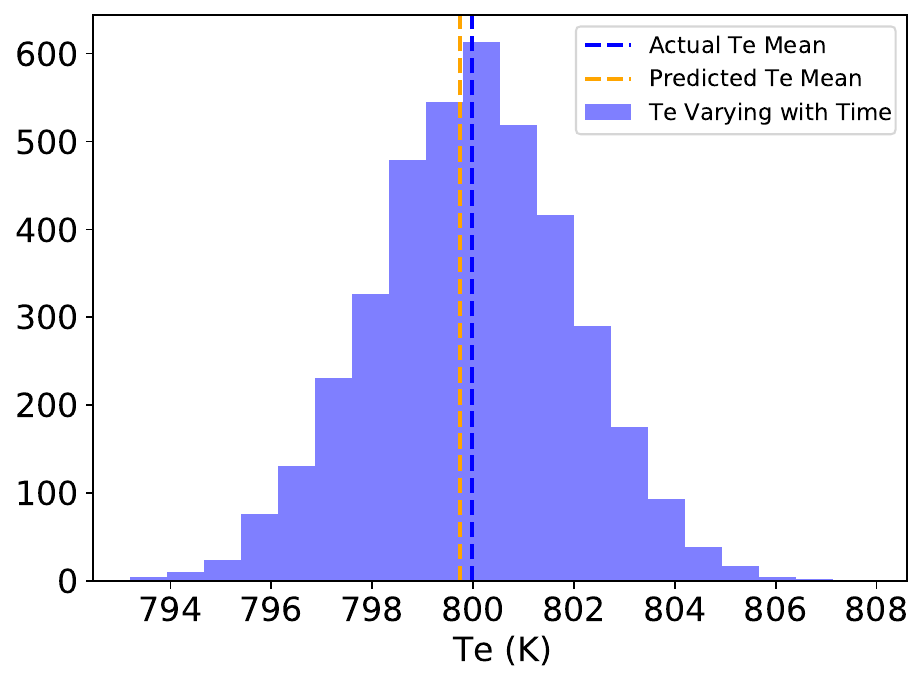}
 \caption{This histogram presents the distribution of Te values containing the entire observation duration in the context of the parametrized signal scenario. The blue dashed line denotes the average value of the actual Te, while the orange dashed line corresponds to the mean of the predicted Te values by the ANN model.}
 \label{fig18}
\end{figure}

\begin{figure}
\centering
 \includegraphics[width=7.2 cm]{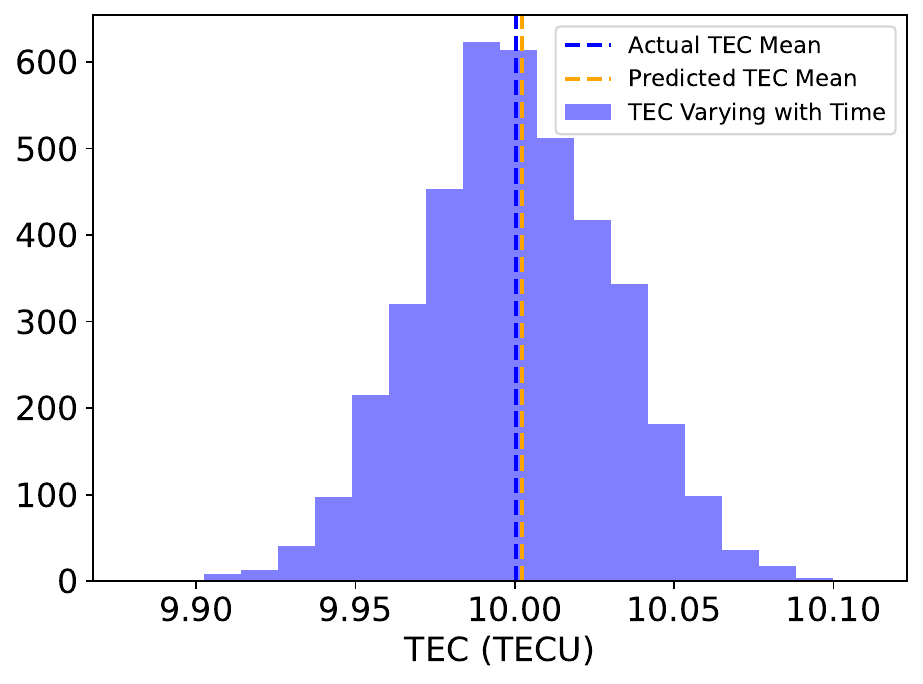}
 \caption{This histogram presents the distribution of TEC values containing the entire observation duration in the context of the Physical signal scenario. The blue dashed line denotes the average value of the actual TEC, while the orange dashed line corresponds to the mean of the predicted TEC values by the ANN model.}
 \label{fig19}
\end{figure}

\begin{figure}
\centering
 \includegraphics[width=7.2 cm]{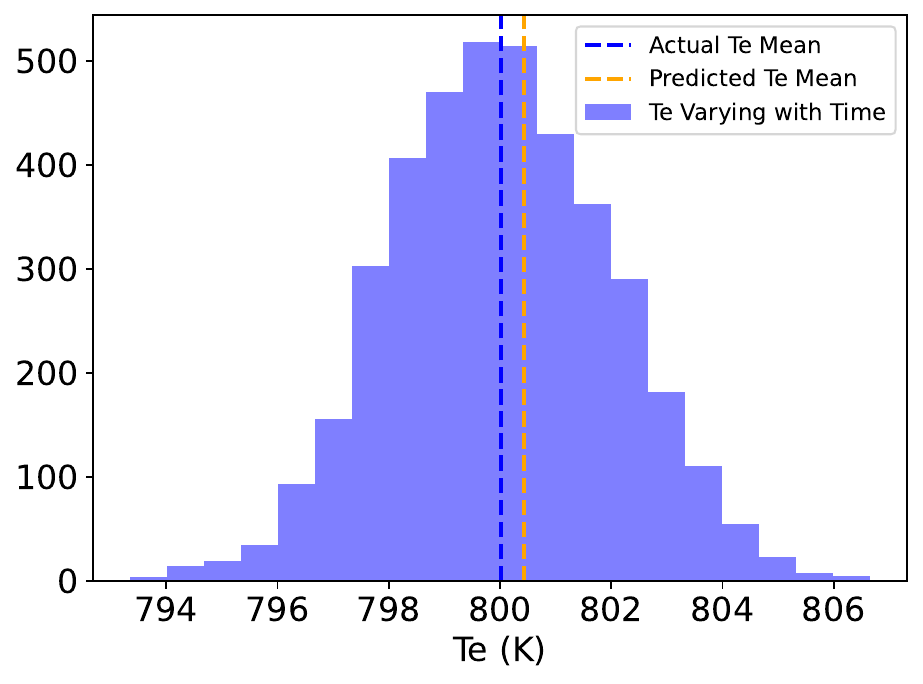}
 \caption{This histogram presents the distribution of Te values containing the entire observation duration in the context of the Physical signal scenario. The blue dashed line denotes the average value of the actual Te, while the orange dashed line corresponds to the mean of the predicted Te values by the ANN model.}
 \label{fig20}
\end{figure}

We conducted further assessments to evaluate the robustness of the ANN model. We used a dynamic ionospheric model with temporal variations. This model assumes random fluctuations in Total Electron Content (TEC) and electron temperature (Te) throughout the observation period. We integrated the F-layer TEC and Te values at 15-minute intervals over a span of 1000 hours to create an observational data set, see Fig. (\ref{fig15}) and Fig. (\ref{fig16}). We calculated the average antenna temperatures measured by the radiometer for each integration over the entire 1000-hour observation period to create the final observation data set that accounts for the time-varying ionospheric effects.
Additionally, we have included the thermal noise associated with the 1000-hour observation in this recorded data set. We feed this noise-added final observation data set to the trained ANN model and extract the parameters. The ANN model demonstrated commendable performance for both parametrized and physically-based signal scenarios. Even in time-varying conditions, the ANN model exhibited accurate predictions. The predicted mean values by the ANN are closely aligned with the actual mean values of TEC and Te, see Fig. (\ref{fig17}, \ref{fig18}, \ref{fig19} and \ref{fig20}) and rest other parameters like signal and foreground are also consistent with the actual one [see Tab. \ref{tab7}, \ref{tab8}]  . This is clearly indicating the ANN model's accuracy and ability to capture complex temporal variations in ionospheric phenomena.

\begin{table*}
\centering
\begin{tabular}{llllllllllllllll}
\hline
Parameters & $\rm J_{ref}$  & $\rm J_{z0 }$   & $\rm X_{z0}$ &$\rm T_{z0}$  & $\rm J_{dz}$ & $\rm X_{dz}$ &$\rm T_{dz}$  &$\rm a_{0}$ &$\rm a_{1}$ & $\rm a_{2}$ & $\rm a_{3}$ & $\rm <TEC>$ & $\rm <T_{e}>$      \\  \hline
Actual Value & 11.690 & 18.540 & 8.680 & 9.770 & 3.310 & 2.820 & 2.830 & 3.323  & -2.354 & -0.0805& 0.0290 & 10.0& 800.00  \\ \hline

Predicted by ANN & 11.823 & 18.693 & 8.627 & 10.144 & 3.407 & 2.901 & 2.853 & 3.336& -2.416& -0.0806 &0.0290 & 10.001& 799.73 \\ \hline

Percentage Error & 1.138 & 0.825 & 0.610 & 3.828 & 2.930 & 2.872 & 0.813 & 0.391 & 2.638 & 0.124 &0.000 & 0.010 & 0.033 \\ \hline

\end{tabular}
\caption{This table presents both the actual parameter values and the values predicted by the ANN, along with the corresponding percentage errors for the time-varying ionospheric model. The parameters include those related to the parametrized signal, foreground, and mean values of TEC and Te.}
\label{tab7}
\end{table*}

\begin{table*}
\centering
\begin{tabular}{lllllllllllll}
\hline
Parameters & $\rm f_{xh}*f_{x}$  &$\rm f_{star }$  & $\rm f_{esc}$ & $\rm N_{\alpha}$ & $\rm a_{0}$ &$\rm a_{1}$ & $\rm a_{2}$ & $\rm a_{3}$ & $\rm <TEC>$ & $\rm <T_{e}>$      \\  \hline
Actual Value & 2.815 & -2.162 & 0.133 & 4.719 & 3.384  & -2.354 & -0.0805& 0.0290 & 10.000& 800.00  \\ \hline

Predicted by ANN & 2.257 & -2.194 & 0.127 & 4.643 & 3.323 & -2.406 & -0.0806 &0.0289 & 10.001& 800.44 \\ \hline

Percentage Error & 19.822 & 1.480 & 4.511 & 1.611 & 1.803 & 2.209 & 0.124 & 0.345 & 0.010 & 0.055 \\ \hline

\end{tabular}
\caption{This table presents both the actual parameter values and the values predicted by the ANN, along with the corresponding percentage errors for the time-varying ionospheric model. The parameters include those related to the physical signal, foreground, and mean values of TEC and Te.}
\label{tab8}
\end{table*}

\section{Summary and Discussions}
\label{ section 10}
In this study, we presented an ANN model to extract the 21-cm global signal by estimating their parameters from a combined spectrum that included signal, foreground, ionospheric effects, and thermal noise. We trained our ANN model with four different scenarios described in detail in Section \ref{ section 7}. To check the robustness of the ANN, we also used two different ways of modelling the global 21-cm signal; one is based on the functional parametrized \citep[][]{mirocha2012optimized, mirocha2015interpreting} and the other one is a physically motivated approach \citep{chatterjee2019ruling, choudhury2021using}. The parameter space in both cases is entirely different; parametrized model parameters are more directly related to IGM properties; however, the physical model includes both IGM and source properties. In the physical model, the parameter $\rm f_{R}$ played the most crucial role in defining the form of the reconstructed signal in the semi-numerical code \citep{choudhury2021using}. A high $\rm f_{R}$ value implies a strong radio background, resulting in a substantial absorption trough signal. In contrast, $\rm f_{R} = 0$ suggests that the excess radio background is turned off, resulting in a conventional signal. In our study, We have taken the traditional data set of the global 21-cm signals from \citep{choudhury2021using}.

For both parametrized and physical models, in the final case studies, Case 1d and Case 2d, the trained ANN model predicted the signal parameters from the test data set with an accuracy of $\ge 96\%$. This clearly demonstrates how a basic ANN model can easily manage up to $13$ parameters ($7$ signal parameters, $4$ foreground parameters, and $2$ ionospheric parameters. We have estimated the uncertainty of the parameter by calculating the RMSE score of the individual parameters for all the cases [ See Tab. \ref{tab4}, \ref{tab6}]. We found the error in the parameter estimation increase when we increase the number of training parameters, e.g., in the first case of the study, when we used signal only (Case 1a), the ANN estimated the parameter with a maximum error of $ \approx 4\%$, but in the final case (Case 1d), the maximum error was $ \approx 6\%$. Similarly, in the second case of the study for Case 2a, the maximum error was $ \approx 3 \%$, but in the final case- Case 2d, the maximum error was $ \approx 5\%$. This clearly indicates that when complexity increased in the training data set, the prediction accuracy slightly decreased with the reference lower complexity training data set [see Tab. \ref{tab3}, \ref{tab4}, \ref{tab5}, and \ref{tab6}]. It means adding complexity to the training data set, making signal extraction more challenging for the network. The accuracy levels will remain high if the network has been trained well enough. We also demonstrated that for the dynamic ionospheric model where TEC and Te are varying randomly, ANN prediction accuracy is still consistent [see Tab. \ref{tab7}, \ref{tab8}]. 

To further emphasise the utility of the ANN method, we conducted an analysis attempting to fit the signal, foreground, and ionospheric effects using a simple analytical approach. This method failed to extract the signal parameters from both scenarios. Conversely, our trained ANN model demonstrated remarkable accuracy in predicting parameters for the same data set, as elaborated in Appendix \ref{A}. Additionally, we evaluated the ANN's performance in terms of accuracy by comparing it with existing prior signal extraction methods. To assess this, we compared the predictions of these prior models with the true parameters, calculating the Mean Absolute Percentage Error (MAPE), as detailed in Appendix \ref{B}. Our findings revealed that the predicted parameters by the ANN model are significantly more accurate than these traditional methods.

The other benefit of using ANN is that it can efficiently extract the observed sky signal's characteristics without modelling and eliminating foreground and ionospheric effects. Compared to the other existing parameter estimation techniques, ANN can extract features from data by building functions that connect the input and output parameters. The ANN model, unlike Bayesian approaches, does not require a defined prior; instead, we must provide training data sets, which may be seen as playing a similar role as the prior in Bayesian techniques. We may avoid computing the likelihood function a large number of times by using ANN to arrive at inferred parameter values. As a result, even when dealing with a larger dimensional parameter space, ANN is computationally faster and more efficient.

In this work, the ANN model is trained for the static as well as time varying ionospheric condition, which is very robust and sensitive for all the given input parameters with their defined parameter ranges used in the preparation of the data sets. By incorporating problems like  beam chromaticity, and other systematic effects, we hope to create a more reliable ANN model in the future. Depending on telescope design and their geomagnetic, various systematics corrupt observations, such as standing waves from cable lengths internal to the system, chromaticity caused by environmental factors like antenna ground planes \citep{2019Singh, 2018Kulkarni}, ionosphere and RFI. These non-astronomical signal need to be modelled for accurate signal extraction. We plan to include these effects in our future study.

\section*{Acknowledgements}
AT would like to thank the Indian Institute of Technology Indore for providing funding for this study in the form of a Teaching Assistantship. AD would like to acknowledge the support from CSIR through EMR-II No. 03(1461)/19. SM and AD acknowledge financial support through the project titled “Observing the Cosmic Dawn in Multicolour using Next Generation Telescopes” funded by the Science and Engineering Research Board (SERB), Department of Science and Technology, Government of India through the Core Research Grant No. CRG/2021/004025.

%%%%%%%%%%%%%%%%%%%%%%%%%%%%%%%%%%%%%%%%%%%%%%%%%%
\section*{Data Availability}
We have used the publicly available code ARES \citep{mirocha2012optimized, mirocha2015interpreting} to simulate the global 21-cm signals for the first case study. For the second case of study, we have used signal data from \citet{choudhury2021using}.

%The inclusion of a Data Availability Statement is a requirement for articles published in MNRAS. Data Availability Statements provide a standardised format for readers to understand the availability of data underlying the research results described in the article. The statement may refer to original data generated in the course of the study or to third-party data analysed in the article. The statement should describe and provide means of access, where possible, by linking to the data or providing the required accession numbers for the relevant databases or DOIs.



\bibliographystyle{mnras}
\bibliography{example} % if your bibtex file is called example.bib

%\begin{thebibliography}{99}
%\end{thebibliography}

%%%%%%%%%%%%%%%%%%%%%%%%%%%%%%%%%%%%%%%%%%%%%%%%%%

%%%%%%%%%%%%%%%%% APPENDICES %%%%%%%%%%%%%%%%%%%%%

\appendix

\section{Extraction of Signal, Foreground and Ionospheric effect Parameters using Analytical Method}
\label{A}
\begin{figure}
\centering
 \includegraphics[width=7.4 cm]{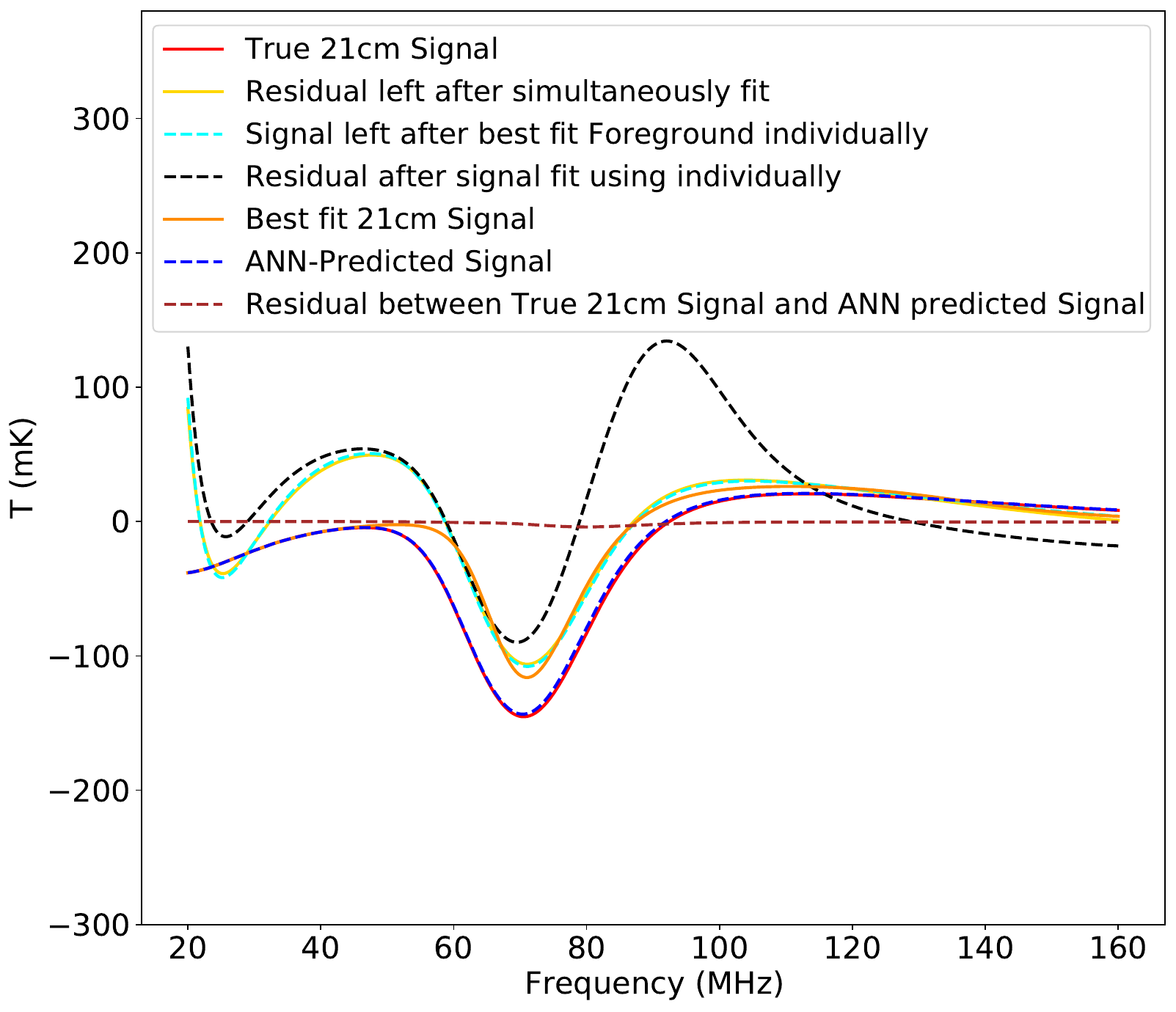}
 \caption{In this figure: true 21-cm signal (solid red line); residual after simultaneous fitting of signal and foreground (yellow solid line); residual after fitting foreground individually (cyan dashed line); best-fitted 21-cm signal from foreground residual (orange solid line); ultimate residual after fitting both foreground and signal individually (black dashed line). Signal reconstructed with ANN predicted parameters (blue dashed line) and residual between the true 21-cm signal and ANN reconstructed signal (brown dashed line).}
 \label{fig21}
\end{figure}

\begin{figure}
\centering
 \includegraphics[width=7.4 cm]{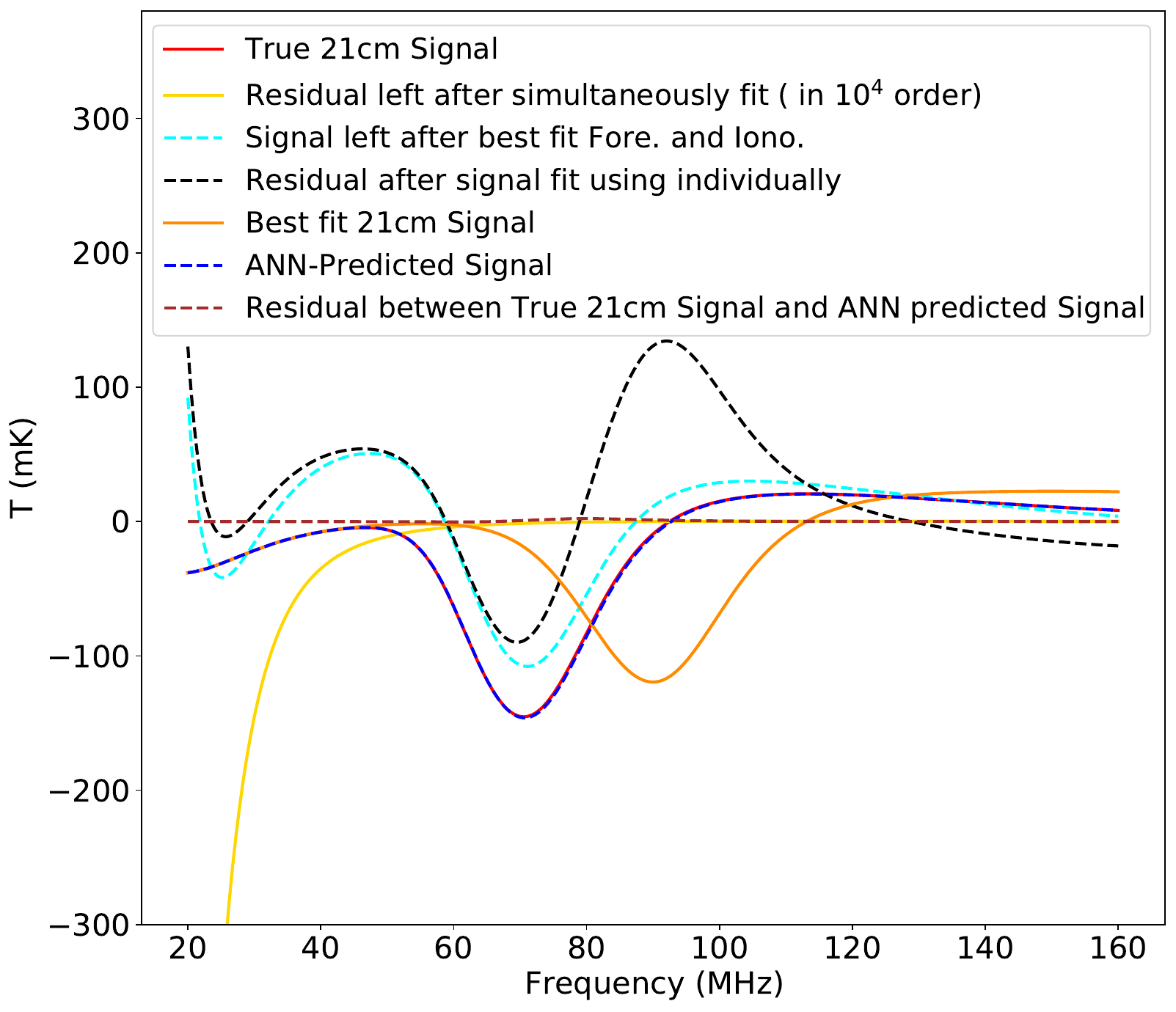}
 \caption{In this figure: true 21-cm signal (solid red line); residual after simultaneous fitting of signal, foreground, and ionospheric effects (yellow solid line); residual after fitting foreground and ionospheric effect individually (cyan dashed line); best-fitted 21-cm signal from foreground residual (orange solid line); ultimate residual after fitting both foreground with ionospheric effects and signal individually (black dashed line). Signal reconstructed with ANN predicted parameters (blue dashed line) and residual between the true 21-cm signal and ANN reconstructed signal (brown dashed line).}
 \label{fig22}
\end{figure}

We attempted to analytically fit two scenarios: one with signal and foreground and another with signal, foreground, and ionospheric effects using Least Square Fit from Scipy libraries in Python. For the first scenario, input data sets are simulated based on equation \ref{eq:20}, and the corresponding true input parameters are listed in Tab. \ref{tab10}. Similarly, for the second scenario, the simulation relied on equation \ref{eq6}, with the true input parameters listed in Tab. \ref{tab11}. We follow two approaches to fit the signal and foreground. In the first approach, We attempted to fit both the signal and foreground simultaneously for the given sky signal simulated using equation \ref{eq:20}, but encountered significant instability in the fit. The residual left after the best-fit signal and foreground is shown in Fig. \ref{fig21}, and best-fit parameters are listed in the table Tab. \ref{tab10}.  
In the second approach, we attempted individual fitting of the foreground and signal from the sky signal, initially fitting the foreground by following equation \ref{eq5} and subtracting its best-fit model from the total observed sky signal. The remaining 21 cm signal was then fitted separately using the signal's parametrized model, described in section \ref{section: 6.1}. However, the fitting function failed to accurately capture the foreground and global signal parameters. It is evident from the distinct residual signal, noticeably different from the input signal, indicating inaccurate foreground fitting (Fig. \ref{fig21}). The small uncertainty in the best-fit foreground parameters indicates under-constraint. The fitting function's performance for signal fitting was notably inadequate. The summarized results are in Tab. \ref{tab10}. We attempted to fit the total sky signal, incorporating the global 21 cm signal, foreground, and ionospheric effects similarly to the previous case. However, the fitting function failed to accurately capture the foreground and ionospheric parameters, as indicated by the statistically significant magnitude of the residual (Fig. \ref{fig22}). We also applied ANN to fit the same data sets for both scenarios, resulting in ANN-predicted parameters that closely approximated the true values of the parameters, see Tab. (\ref{tab10}, \ref{tab11}) and reconstructed 21-cm signals in Fig. \ref{fig21} and \ref{fig22}, accompanied by residuals from the true input signals.

\begin{table}
\setlength{\tabcolsep}{2.5 pt}
\begin{tabular}{|lllll|} \\ \hline
\multicolumn{1}{|l|}{Parameters} & \multicolumn{1}{l|}{True Value} & \multicolumn{2}{l|}{\begin{tabular}[c]{@{}l@{}}Analytical Method\\        (Best Fit Values)\end{tabular}} & ANN        \\ \hline
\multicolumn{1}{|l|}{}           & \multicolumn{1}{l|}{}           & \multicolumn{1}{l|}{Simultaneously}            & \multicolumn{1}{l|}{Individually}                        & Prediction \\ \\ \hline

$\rm J_{ref}$   & \bf{11.6900}    &  -5.6409$\times 10^{5}$  &(0.0036  $\pm$ 8.7395)$\times 10^{7}$   & 11.5351        \\
$\rm J_{z0 }$   & \bf{18.5400}    &  5.8956$\times 10^{3}$   & 9.1845  $\pm$ 2.7466$\times 10^{3}$       & 18.5470          \\
$\rm X_{z0}$    &\bf{ 8.6800}     &  7.9235$\times 10^{5}$   & 9.1432  $\pm$ 0.2201                   & 8.5926          \\
$\rm T_{z0}$    & \bf{9.7700}     &  6.3436$\times 10^{5}$   & 10.4985 $\pm$ 0.3743          & 9.8942          \\
$\rm J_{dz}$    & \bf{3.3100}     &  6.3949$\times 10^{4}$   & 2.2116 $\pm$ 0.2146            & 3.3012          \\
$\rm X_{dz}$    & \bf{2.8200}     & -7.9105e$\times 10^{4}$  & 2.9636 $\pm$ 0.4439         & 2.8133          \\
$\rm T_{dz}$    & \bf{2.8300}     &  6.7207$\times 10^{2}$   & 1.5002 $\pm$ 0.3700     & 2.8163          \\
$\rm a_{0}$     & \bf{3.30942}    &  3.30940    & 3.30941  $\pm$ 3.478 $\times 10^{-7}$        & 3.31261          \\
$\rm a_{1}$     & \bf{-2.40960}   &  -2.40963   & -2.40959 $\pm $2.567 $\times 10^{-6}$        & -2.41448 \\
$\rm a_{2}$     & \bf{-0.08062}   &  -0.08059   & -0.08054 $\pm$ 9.015e $\times 10^{-6}$        & -0.08063 \\
$\rm a_{3}$     & \bf{ 0.02898}   &  0.02905    &  0.02909 $\pm$ 9.529e $\times 10^{-6}$      & 0.02897           \\\hline
     
\end{tabular}

\caption{This table contains the true signal and foreground parameter values and their best-fit estimates and uncertainties obtained through a simple analytical method. Additionally, it includes parameter values predicted by the ANN model.}
\label{tab10}
\end{table}

\begin{table}
\setlength{\tabcolsep}{1.0 pt}
\begin{tabular}{|lllll|} \\ \hline
\multicolumn{1}{|l|}{Parameters} & \multicolumn{1}{l|}{True Value} & \multicolumn{2}{l|}{\begin{tabular}[c]{@{}l@{}}Analytical Method\\        (Best Fit Values)\end{tabular}} & ANN        \\ \hline
\multicolumn{1}{|l|}{}           & \multicolumn{1}{l|}{}           & \multicolumn{1}{l|}{Simultaneously}                  & \multicolumn{1}{l|}{Individually}                  & Prediction \\ \hline

$\rm J_{ref}$   & \bf{11.6900}   & -5.9815 $\times 10^{2}$    &  10.9367 $\pm$ 9.49 $\times 10^{5}$   & 11.6877  \\
$\rm J_{z0 }$   & \bf{18.5400}   &  1.0756 $\times 10^{2}$    &  12.1768 $\pm$ 1.58 $\times 10^{5}$   & 18.5304  \\
$\rm X_{z0}$    &\bf{ 8.6800}    & -5.5549 $\times 10^{6}$    &  7.170 $\pm$ 4.91 $\times 10^{5}$     & 8.6748   \\
$\rm T_{z0}$    & \bf{9.7700}    &  9.7794 $\times 10^{4}$    &  8.083 $\pm$ 2.82 $\times 10^{4}$     & 9.7236   \\
$\rm J_{dz}$    & \bf{3.3100}    &  -4.1476                   &  3.2672 $\pm$ 1.83 $\times 10^{4}$    & 3.3036   \\
$\rm X_{dz}$    & \bf{2.8200}    &  2.38 $\times 10^{5}$      &  2.0259 $\pm$ 3.12 $\times 10^{5}$    & 2.8140   \\
$\rm T_{dz}$    & \bf{2.8300}    &  4.33 $\times 10^{3}$      &  0.3418 $\pm$ 1.43 $\times 10^{4}$    & 2.8284   \\
$\rm a_{0}$     & \bf{3.30942}   &  3.30932                   & 3.30930  $\pm$ 5.98 $\times 10^{-6}$  & 3.30651 \\
$\rm a_{1}$     & \bf{-2.40960}  & -2.40946                   & -2.40951  $\pm$ 7.57 $\times 10^{-6}$ &-2.40808  \\
$\rm a_{2}$     & \bf{-0.08062}  & -0.08074                   & -0.08071  $\pm$ 3.48 $\times 10^{-5}$ & 0.08063  \\
$\rm a_{3}$     & \bf{ 0.02898}  &  0.03076                   & 0.03079  $\pm$ 8.36 $\times 10^{-5}$  & 0.02897    \\
$\rm TEC$       & \bf{ 10.0000}  &  1.0299                    & 10.1183  $\pm$ 0.0076                 & 9.9998      \\ 
$\rm T_{e}$     & \bf{ 800.0000} &  275.6359                 & 813.7119  $\pm$ 0.8743                 & 799.8309      \\ \hline     
\end{tabular}

\caption{This table contains the true signal, foreground and Ionospheric parameter values and their best-fit estimates and uncertainties obtained through a simple analytical method. Additionally, it includes parameter values predicted by the ANN model.}
\label{tab11}
\end{table}

\section{Comparison with Other Existing Techniques}
\label{B}
We evaluated the accuracy of our ANN predictions in comparison to other methods by calculating the Mean Absolute Percentage Error (MAPE) detailed in Tab. \ref{tab12}. Our study demonstrated that while some prior techniques performed well with simpler signal models (those with fewer free parameters), they faltered when dealing with more complex signal models requiring additional parameters. In contrast, the ANN model consistently outperformed these methods. Tab. \ref{tab12} demonstrates that our ANN predictions exhibited less than 5\% error across all parameters, regardless of the signal scenarios, including foreground with ionospheric effects. For the physical signals, except for one parameter with more than 10\% error, all other parameters are accurately constrained, with most falling below the range of 5\% to 1.0\%  error.

\begin{table*}
\setlength{\tabcolsep}{0.01 pt}
\begin{tabular}{cll}
\hline
\textbf{Methods}                                                                                                                                                                        & \multicolumn{1}{c}{\textbf{\begin{tabular}[C]{@{}c@{}}data sets and Associated Parameters\end{tabular}}}                                                                          & \multicolumn{1}{c}{\textbf{MAPE}}   \\ \hline
\begin{tabular}[c]{@{}c@{}}MCMC \\ \citep{harker2012mcmc}\end{tabular}                                                                                                                 & \begin{tabular}[c]{@{}l@{}}Signal only  (Turning Point Model)\\ \hline Parameters : ($\rm \nu_{B}$, $ \rm T_{B}$, $\rm \nu_{C}$, $T_{C}$, $\nu_{D}$, $T_{D}$)\end{tabular}                                                      & \begin{tabular}[c]{@{}l@{}}$\rm \nu_{B}$: 3.77,  $\rm T_{B}$: \textbf{591.43}, $\rm \nu_{C}$: 0.50 \\ $\rm T_{C}$: 12.28, $\rm \nu_{D}$: 0.22,    $\rm T_{D}$: \textbf{39.15}\end{tabular}                                                                                                                          \\ \hline
\begin{tabular}[c]{@{}c@{}} Singular Value \\ Decomposition (SVD) with MCMC \\ \citep{Rapetti_2020}\end{tabular}                                                                                                      & \begin{tabular}[c]{@{}l@{}}Signal with Beam Weighted Foreground \\ \hline Signal: Flattened Gaussian Model \\ Parameters : ( $\rm A$, $ \rm \nu_{0}$, $\rm w$, $\rm \tau$) \\ \hline
Signal : Turning Point Model \\ Parameters : ($\rm \nu_{A}$, $ \rm T_{A}$, $\rm \nu_{B}$, $ \rm T_{B}$, $\rm \nu_{C}$, $T_{C}$, $\nu_{D}$, $T_{D}$)  \end{tabular}                                     & \begin{tabular}[c]{@{}l@{}}Flattened Gaussian  Model\\ $\rm A$: 0.050, $\rm \nu_{0}$: 0.003 $\rm w$: 0.052,  $\rm \tau$: 0.683 \\ \hline Turning Point Model \\ $\rm \nu_{A}$: 14.19, $ \rm T_{A}$: \textbf{42.01}, $\rm \nu_{B}$: 0.52, $ \rm T_{B}$: \textbf{257.22}  \\ $\rm \nu_{C}$: 6.91, $ \rm T_{C}$: 0.85, $\rm \nu_{D}$: 0.10, $ \rm T_{D}$: 10.66, $\rm \nu_{E}$: 1.75\end{tabular} \\ \hline
\begin{tabular}[c]{@{}c@{}}SVD with MCMC\\ \citep{Tauscher2021}\end{tabular}                                                                                                                & \begin{tabular}[c]{@{}l@{}}Signal and Beam Weighted Foreground with \\ the Instrument effect\\ \hline Signal: Turning Point model \\ Parameters: ($\rm \nu_{B}$, $ \rm T_{B}$, $\rm \nu_{C}$, $T_{C}$, $\nu_{D}$, $T_{D}$)\end{tabular}  & \begin{tabular}[c]{@{}l@{}}Turning Point Model\\ $\rm \nu_{B}$: 0.011, $ \rm T_{B}$: \textbf{Not constrained},  $\rm \nu_{C}$: 0.754, \\ $ \rm T_{C}$: 1.942, $\rm \nu_{D}$: 0.278, $ \rm T_{D}$: \textbf{Not constrained}\end{tabular}     \\ \hline

\multicolumn{1}{c}{\begin{tabular}[c]{@{}l@{}}Least Square Fit using Scipy\\ \citep{emaa_2021}\end{tabular}}                                                                            & \begin{tabular}[c]{@{}l@{}}Signal, Foreground with Ionospheric effects\\ \hline Signal : Normal Gaussian Signal \\   Parameters : ($\rm A$ , $\rm \mu$ , $\rm \sigma$)\end{tabular}                      & \begin{tabular}[c]{@{}l@{}} F-layer snapshot : $\rm A$: 18.87,   $\rm \mu$: 0.77, $\rm \sigma$: 6.95\\ \hline D-layer snapshot : $\rm A$: \textbf{24.53},  $\rm \mu$: 5.69,  $\rm \sigma$: \textbf{66.85}\end{tabular}    \\ \hline
\multicolumn{1}{c}{\begin{tabular}[c]{@{}l@{}}1) Maximally Smooth Function (MSF)\\ 2) Partial Smooth Function (PSF)\\ 3) Polynomial Fit \\ \citep{Bevins_2021}\end{tabular}} & \begin{tabular}[c]{@{}l@{}}Signal with Foreground   \\ \hline Signal: Simulated Global 21-cm \\  Signals by \citep{cohen2017charting}\end{tabular}                                    & \begin{tabular}[c]{@{}l@{}}No of the Turning points predicted in residual\\ 1) MSF: 56\\ 2) PSF: 99\\ 3) Polynomial Fit (N=5): 100\end{tabular} \\ \hline
\multicolumn{1}{c}{\begin{tabular}[c]{@{}l@{}} In this work using  ANN \end{tabular}} & \begin{tabular}[c]{@{}l@{}}Signal, Foreground with Ionospheric effects\\ \hline 1) Signal:  parametrized model \\ Parameters : ($\rm J_{ref}$, $\rm J_{z0 }$, $\rm X_{z0}$, $\rm T_{z0}$, $\rm J_{dz}$, $\rm X_{dz}$, $\rm T_{dz}$, \\ $\rm a_{0}$, $\rm a_{1}$, $\rm a_{2}$, $\rm a_{3}$, $\rm TEC$, $\rm T_{e}$)    \\ \hline 2) Signal : Physical model \\
Parameters: ($\rm f_{xh}*f_{x}$, $\rm f_{star }$, $\rm f_{esc}$, $\rm N_{\alpha}$,  $\rm a_{0}$, $\rm a_{1}$,  $\rm a_{2}$,  $\rm a_{3}$,  $\rm TEC$, $\rm T_{e}$) \end{tabular} & \begin{tabular}[c]{@{}l@{}} 1) parametrized Model \\   $\rm J_{ref}$: 4.47, $\rm J_{z0 }$: 4.46,  $\rm X_{z0}$: 4.34, $\rm T_{z0}$: 4.11, $\rm J_{dz}$: 4.36, \\ $\rm X_{dz}$: 4.07, $\rm T_{dz}$: 4.15, $\rm a_{0}$: 1.05, $\rm a_{1}$: 0.26, $\rm a_{2}$: 0.08, \\ $\rm a_{3}$: 0.07,  $\rm TEC$: 0.07, $\rm T_{e}: 0.07$
\\ \hline 2) Physical Model \\ $\rm f_{xh}*f_{x}$: 13.31, $\rm f_{star }$: 0.86,  $\rm f_{esc}$: 4.91, $\rm N_{\alpha}$: 0.71, $\rm a_{0}$: 1.11, \\ $\rm a_{1}$: 0.33,  $\rm a_{2}$: 0.07, $\rm a_{3}: 0.07 $, $\rm TEC$: 0.08, $\rm T_{e}$: 0.07  \end{tabular} \\\hline
\end{tabular}
\caption{This table presents an overview of methodologies and signal models from prior studies used for global 21-cm signal parameter extraction. We calculated the Mean Absolute Percentage Error (MAPE) between predicted and true parameters using these methods and compared the results with predictions made by ANN in our study. Parameters with MAPE values $> 20$ or unconstrained are highlighted in bold font.}
\label{tab12}
\end{table*}

%%%%%%%%%%%%%%%%%%%%%%%%%%%%%%%%%%%%%%%%%%%%%%%%%%

% Don't change these lines
\bsp	% typesetting comment
\label{lastpage}
\end{document}